\pdfoutput=1
\documentclass[sigconf,nonacm]{acmart}

\usepackage{algorithm}
\usepackage{algpseudocode}
\usepackage{amsmath}
\usepackage{array}
\usepackage{mathtools}
\usepackage{subcaption}
\usepackage{tikz}
\usetikzlibrary{calc, positioning}
\usepackage{pgfplots}
\pgfplotsset{compat=1.18}
\usepackage{subcaption}
\usepackage{makecell}
\usepackage{stfloats}

\usetikzlibrary{positioning,calc,shapes.geometric,decorations.pathreplacing}

\newcolumntype{P}[1]{>{\centering\arraybackslash}p{#1}}

\makeatletter
\newcommand{\thickhline}{%
    \noalign {\ifnum 0=`}\fi \hrule height 1pt
    \futurelet \reserved@a \@xhline
}
\newcolumntype{"}{@{\hskip\tabcolsep\vrule width 1pt\hskip\tabcolsep}}
\makeatother

\newcommand{\mytilde}{\raisebox{0.5ex}{\texttildelow}}

\newlength{\DepthReference}
\newlength{\HeightReference}
\newlength{\Width}%
\newcommand{\MyColorBox}[2][red]%
{%
    \settowidth{\Width}{#2}%
    \colorbox{#1}%
    {%
        \raisebox{-\DepthReference}%
        {%
                \parbox[b][\HeightReference+\DepthReference][c]{\Width}{\centering#2}%
        }%
    }%
}

\algrenewcommand\algorithmicrequire{\textbf{Input:}}
\algrenewcommand\algorithmicensure{\textbf{Output:}}

\algblock{ParFor}{EndParFor}
\algnewcommand\algorithmicparfor{\textbf{parallel for}}
\algnewcommand\algorithmicpardo{\textbf{do}}
\algnewcommand\algorithmicendparfor{\textbf{end\ for}}
\algrenewtext{ParFor}[1]{\algorithmicparfor\ #1\ \algorithmicpardo}
\algrenewtext{EndParFor}{\algorithmicendparfor}

\definecolor{supergreen}{RGB}{0,170,0}

\DeclareMathOperator{\Z}{\mathbb{Z}}
\DeclareMathOperator{\R}{\mathbb{R}}
\DeclareMathOperator{\C}{\mathbb{C}}

\newcommand{\norm}[1]{\left\lVert#1\right\rVert}

\DeclareMathOperator*{\argmin}{argmin}
\newcommand{\gaussian}{\mathcal{N}}

\DeclarePairedDelimiter{\ceil}{\lceil}{\rceil}
\DeclarePairedDelimiter{\floor}{\lfloor}{\rfloor}

\DeclareMathOperator{\cmp}{cmp}

\DeclareMathOperator{\sign}{sign}

\newcommand{\enc}{\textsf{Enc}}
\newcommand{\dec}{\textsf{Dec}}

\newcommand{\maskR}{\textsf{MaskR}}
\newcommand{\maskC}{\textsf{MaskC}}
\newcommand{\sumR}{\textsf{SumR}}
\newcommand{\sumC}{\textsf{SumC}}
\newcommand{\replR}{\textsf{ReplR}}
\newcommand{\replC}{\textsf{ReplC}}
\newcommand{\transR}{\textsf{TransR}}
\newcommand{\transC}{\textsf{TransC}}

\newcommand{\vectorLength}{M}
\renewcommand{\vector}{v}
\newcommand{\encVector}{V}

\newcommand{\encComparison}{C}
\newcommand{\ranking}{r}
\newcommand{\encRanking}{R}

\newcommand{\encArgminVar}{A}

\newcommand{\tabcolsepValue}{1.2pt}
\newcommand{\arraystretchValue}{1.3}

\newcommand{\repository}{https://anonymous.4open.science/r/secure-vertical-kmeans-EB62}
\newcommand{\point}{x}
\newcommand{\encPoint}{X}
\newcommand{\numClusters}{k}
\newcommand{\cluster}{C}
\newcommand{\centroid}{c}
\newcommand{\numParties}{N}
\newcommand{\numPoints}{n}
\newcommand{\ringDim}{n}
\newcommand{\numDims}{d}
\newcommand{\distance}{d}
\newcommand{\encDistance}{D}
\newcommand{\valueBound}{B}
\newcommand{\clusterSize}{t}
\newcommand{\encClusterSize}{T}
\newcommand{\sumPoints}{s}
\newcommand{\encSumPoints}{S}
\newcommand{\numRounds}{r}
\newcommand{\encReplResult}{R}

\newcommand{\groundTruth}{l}

\title[Privacy-Preserving Vertical K-Means Clustering]{Privacy-Preserving Vertical K-Means Clustering}

\author{Federico Mazzone}
\authornote{This work was done during a visiting period at the University of Waterloo (Canada).}
\email{f.mazzone@utwente.nl}
\affiliation{%
  \institution{University of Twente}
  \city{Enschede}
  \country{The Netherlands}
}

\author{Trevor Brown}
\email{trevor.brown@uwaterloo.ca}
\affiliation{%
  \institution{University of Waterloo}
  \city{Waterloo}
  \country{Canada}
}

\author{Florian Kerschbaum}
\email{florian.kerschbaum@uwaterloo.ca}
\affiliation{%
  \institution{University of Waterloo}
  \city{Waterloo}
  \country{Canada}
}

\author{Kevin H.~Wilson}
\email{kevin.h.wilson@borealisai.com}
\affiliation{%
  \institution{RBC Borealis}
  \city{Toronto}
  \country{Canada}
}

\author{Maarten Everts}
\authornote{Also affiliated with Linksight.}
\email{maarten.everts@utwente.nl}
\affiliation{%
  \institution{University of Twente}
  \city{Enschede}
  \country{The Netherlands}
}

\author{Florian Hahn}
\email{f.w.hahn@utwente.nl}
\affiliation{%
  \institution{University of Twente}
  \city{Enschede}
  \country{The Netherlands}
}

\author{Andreas Peter}
\email{andreas.peter@uni-oldenburg.de}
\affiliation{%
  \institution{University of Oldenburg}
  \city{Oldenburg}
  \country{Germany}
}

\renewcommand{\shortauthors}{Mazzone et al.}

\keywords{K-Means Clustering, Homomorphic Encryption, Differential Privacy, Vertically Partitioned Data, Lloyd's Algorithm, Secure Data Analysis}

\begin{document}

\begin{abstract}
Clustering is a fundamental data processing task used for grouping records based on one or more features.
In the vertically partitioned setting, data is distributed among entities (e.g., hospitals, insurers, or government agencies), with each holding only a subset of those features.
A key challenge in this scenario is that computing distances between records requires access to all distributed features, which may be privacy-sensitive and cannot be directly shared with other parties.
The goal is to compute the joint clusters while preserving the privacy of each entity’s dataset.

Existing solutions using secret sharing or garbled circuits implement privacy-preserving variants of Lloyd's algorithm but incur high communication costs, scaling as $O(nkt)$, where $n$ is the number of data points, $k$ the number of clusters, and $t$ the number of rounds.
These methods become impractical for large datasets or several parties, limiting their use to LAN settings only.
On the other hand, a different line of solutions rely on differential privacy (DP) to outsource the local features of the parties to a central server.
However, they often significantly degrade the utility of the clustering outcome due to excessive noise.

In this work, we propose a novel solution based on homomorphic encryption and DP, reducing communication complexity to $O(n+kt)$.
In our method, parties securely outsource their features once, allowing a computing party to perform clustering operations under encryption.
DP is applied only to the clusters' centroids, ensuring privacy with minimal impact on utility.
We show the efficiency and scalability of our solution by assessing it on a variety of real-world and synthetic datasets.
For example, clustering 100,000 two-dimensional points into five clusters requires only 73MB of communication, compared to 101GB for existing works, and completes in just under 3 minutes on a 100Mbps network, whereas existing works take over 1 day.
This makes our solution practical even for WAN deployments, all while maintaining accuracy comparable to plaintext k-means algorithms.
\end{abstract}

\maketitle

\section{Introduction}

Collaborative machine learning enables multiple parties to jointly analyze and process data while keeping their individual datasets private.
In the \textit{vertically-partitioned} (VP) setting, different entities possess complementary features of the same set of data points.
For instance, in healthcare, multiple hospitals may each have patient records with different types of medical data (e.g., one hospital may have diagnostic images or lab results, while another holds treatment history or medication prescriptions).
Similarly, in financial services, different institutions may have access to different aspects of customer data, such as spending habits or loan histories.
Another example is customer modeling, where a company may have demographic data while a partner company has transactional data.
In such scenarios, the entities wish to collaboratively analyze the data to gain insights while preserving the privacy of their individual datasets.

A common machine learning task in such settings is clustering, where data points are grouped based on their similarities.
Clustering, particularly k-means clustering, is an unsupervised learning technique aimed at partitioning a set of $n$ data points into $k$ clusters, where each data point is assigned to the cluster with the closest mean point, called \textit{centroid}.
The most widely used algorithm for solving this problem is Lloyd’s algorithm~\cite{lloyd1982least}, an iterative method that runs in polynomial time by refining centroids until convergence.
In the VP setting, clustering has numerous applications, such as segmenting customers for personalized offers, identifying subgroups of patients for treatment analysis, or clustering financial profiles for risk assessment --- all without compromising the privacy of each entity's data~\cite{chaturvedi1997feature,elmisery2010privacy,vaidya2003privacy,punj1983cluster}.

Clustering in the vertical setting presents many challenges.
The main difficulty arises from the fact that each party holds only a subset of the data features, hindering a direct computation of distances between centroids and data points.
Several works in the literature have addressed this problem in a privacy-preserving way.
These works can be mainly divided in two categories:
\begin{enumerate}
    \item \textbf{Multiparty computation (MPC)-based solutions}, which use cryptographic techniques like garbled circuits, secret sharing, and oblivious transfer to securely compute the clusters~\cite{vaidya2003privacy, jagannathan2005privacy, bunn2007secure, mohassel2020practical}.
    \item \textbf{Differential privacy (DP)-based solutions}, which add noise directly to the data or some encoding of it to preserve privacy while outsourcing computations~\cite{su2016differentially, su2017differentially, balcan2017differentially, li2022differentially}.
\end{enumerate}

However, both approaches have significant limitations.
MPC-based protocols incur prohibitively high communication costs, which prevents these solutions from scaling well with the dataset size or the number of parties involved.
The most efficient work in this category~\cite{mohassel2020practical} has a communication complexity of  $O(nkt)$, where $t$ is the number of iterations of Lloyd's algorithm.
This results in gigabytes of data to be exchanged, even for relatively small datasets.
For example, clustering 10,000, 100,000, and 1,000,000 points in five clusters requires 10GB, 101GB, and 1TB of communication, respectively.
Such high costs make MPC-based approaches impractical for large datasets or wide-area network (WAN) settings, where bandwidth is limited.
Furthermore, MPC-based solutions provide no privacy guarantees for the final clustering output, which could leak information about the input data.
On the other hand, DP-based solutions provide a faster alternative but often sacrifice accuracy.
The noise added directly to the input data can significantly degrade clustering utility, especially for high-dimensional data or large numbers of clusters.
Even the best DP-solution to date~\cite{li2022differentially} suffers from an accuracy loss of up to 30 percentage points compared to the plaintext version of the algorithm, despite using a moderate privacy budget ($\epsilon = 1$).

In this paper, we propose a new protocol for clustering in the VP setting that overcomes these limitations.
Our approach employs CKKS~\cite{cheon2017homomorphic}, an homomorphic encryption (HE) scheme, to encrypt and securely outsource the input data upfront.
A computing party then executes Lloyd’s algorithm on the encrypted data, using an efficient re-encoding technique to exploit the Single-Instruction-Multiple-Data (SIMD) capabilities of CKKS, enabling highly-parallelized operations under encryption and a low computation time --- traditionally a bottleneck in HE applications.
After each iteration, a differentially private version of the centroids is disclosed to refresh the ciphertext noise, following the same approach of the DP Lloyd's algorithm for the central model~\cite{blum2005practical,su2016differentially}.
This mechanism reduces the communication complexity to $O(n+kt)$ while preserving clustering accuracy, as DP is applied exclusively to the centroids rather than the data points.

Our solution is practical for clustering datasets with millions of points in just a few minutes.
Compared to the state-of-the-art MPC solution~\cite{mohassel2020practical}, our method reduces the communication size by up to three orders of magnitude and runtime by up to four orders of magnitude in WAN settings.
For instance, for five clusters, our protocol requires just 19MB (526x less), 73MB (1384x less), and 563MB (1794x less) of communication for 10,000, 100,000, and 1,000,000 points, respectively.
The clustering completes in 1.70 minutes, 2.51 minutes, and 22.21 minutes, respectively, dramatically faster than MPC-based solutions, which can take hours or days depending on network conditions.
Moreover, our protocol achieves clustering accuracy comparable to plaintext computation, significantly outperforming existing DP-based solutions in utility.

Our contributions are as follows:
\begin{itemize}
    \item A novel protocol for privacy-preserving k-means clustering in the VP setting, scalable to large datasets and efficient even in constrained network environments.
    \item An optimized method for packing multiple argmin computations in a single CKKS ciphertext.
    \item A thorough experimental evaluation on a variety of real-world and synthetic datasets across different network configurations.
\end{itemize}


\section{Background}

\subsection{Lloyd's Algorithm}

K-Means clustering is a problem of partitioning $\numPoints$ points $\point_1, \dots, \point_\numPoints \in \R^\numDims$ into $\numClusters$ clusters $\cluster_1, \dots, \cluster_\numClusters$ such that the total squared distance between the points and their corresponding cluster centers are the least.
Namely, we look for a partition that minimizes the quantity
\begin{equation*}
    \sum_{j = 1}^{\numClusters}{\sum_{\point_i \in \cluster_j}{\norm{\point_i - \centroid_j}^2}}
\end{equation*}
where $\centroid_j := \sum_{\point_i \in \cluster_j}{\point_i} \,/\, | \cluster_j|$ is the mean or \textit{centroid} of cluster $\cluster_j$.

The k-means problem is NP-hard, but local optimization algorithms can find a sub-optimal solution in polynomial time.
The most widely used such algorithm is Lloyd's algorithm~\cite{lloyd1982least}.
Given an initial set of centroids $\centroid_1, \dots, \centroid_\numClusters$, the algorithm iteratively performs two steps:
\begin{enumerate}
    \item Centers-to-Clusters: compute the distance $\distance_{ij} = \norm{\point_i - \centroid_j}$ between each point $\point_i$ and centroid $\centroid_j$, and assign point $\point_i$ to the cluster of the closest centroid, that is
    $$\cluster_j = \Bigl\{\point_i : j = \argmin_{\centroid_1, \dots, \centroid_\numClusters}{\distance_{ij}}\Bigr\} \enspace .$$
    \item Clusters-to-Centers: update each centroid to be the mean of the points in the corresponding cluster:
    $$\centroid_j = \frac{1}{|\cluster_j|} \sum_{\point_i \in \cluster_j}{\point_i} \enspace .$$
\end{enumerate}
These two steps are repeated until convergence.


\subsection{Differential Privacy}
\label{sec:background:dp}

Differential Privacy~\cite{dwork2006calibrating} is a privacy-preserving technique that aims to conceal the presence or absence of individual records in a dataset during data analysis.
In particular, given a privacy budget $\epsilon$ and a failure probability $\delta$, a randomized algorithm $f : A \to B$ is $(\epsilon, \delta)$-DP if for any pair of datasets $D,D' \in A$ that differ in exactly one record
\begin{equation*}
    \Pr[f(D) \in O] \le e^\epsilon \Pr[f(D') \in O] + \delta, \qquad \forall O \subseteq B \enspace .
\end{equation*}
DP usually works by injecting precisely sampled noise during the algorithm computations or directly into its input data.

\paragraph{Gaussian Mechanism}
Given a real function $f : A \to \R$, we know that $F(x) := f(x) + \gaussian(0, \sigma^2)$ is $(\epsilon,\delta)$-DP if
$$\sigma = \sqrt{2 \log(1.25 / \delta)} \, \frac{s}{\epsilon}$$
where $s = \max_{D\sim D'}{|f(D) - f(D')|}$ is the \textit{sensitivity} of $f$, which measures the maximum variation of the algorithm's output when exactly one input record is modified~\cite{dwork2014algorithmic}.

\paragraph{Composition Theorems}
Given a function $f$ that is $(\epsilon, \delta)$-DP, we know that its $\numRounds$-fold sequential composition $f^\numRounds$ is also DP.
In particular, according to the simple composition theorem, we have that $f^\numRounds$ is $(\epsilon',\delta')$-DP with $\epsilon' = \numRounds \epsilon$ and $\delta' = \numRounds \delta$.
While, according to the advanced composition theorem, we have that for any $\delta' > 0$, $f^\numRounds$ is $(\epsilon', \numRounds \delta + \delta')$-DP with $\epsilon' = 2 \epsilon \sqrt{2 \numRounds \log(1 / \delta')}$~\cite{dwork2014algorithmic}.
Having these composition theorems is especially useful when handling iterative algorithms --- like in our case --- to understand how much privacy budget to allocate to each iteration.

\subsection{Homomorphic Encryption and Argmin}
\label{sec:background:he}

Homomorphic Encryption is a cryptographic primitive that enables performing operations on encrypted data, without decrypting them first.
CKKS~\cite{cheon2017homomorphic}, in particular, is a fully HE scheme based on the RLWE problem.
It works with residual polynomial rings of the form $R_q = \Z_q[x] / (x^\ringDim + 1)$, where the ring dimension $\ringDim$ is a power of two.
Messages from $\C^{\ringDim/2}$ are encoded into plaintexts, which can embed vectors of up to $\ringDim / 2$ slots.
The scheme operates with floating-point values and it is intrinsically approximate.
CKKS natively supports three homomorphic operations on ciphertexts: (1) component-wise addition $(X+Y)$, (2) component-wise multiplication $(X \cdot Y)$, and (3) vector rotation (left $X \ll r$, and right $X \gg r$).
The component-wise operations allow processing many inputs concurrently, which makes CKKS suitable for Single-Instruction-Multiple-Data (SIMD).
By combining additions and multiplications it is possible to evaluate any polynomial, while for non-polynomial functions the usual solution consists of approximating the function with a high-degree polynomial.
For instance, to evaluate the comparison function, we can approximate the sign function
\begin{equation*}
\sign(z) =
\begin{cases}
    1 & \text{if } z > 0 \\
    0 & \text{if } z = 0 \\
    -1 & \text{if } z < 0
\end{cases}
\end{equation*}
using Chebyshev interpolation, and compare any two values $x, y$ by defining $\cmp(x,y) := \sign(x - y) / 2 + 0.5 \in \{0, 0.5, 1\}$, which tells you whether $x > y$.

A fundamental step in Lloyd's algorithm is finding the closest centroid to each data point, which in our case translates to computing the argmin of $\numClusters$ values under encryption.
To do so, we employ the argmin approach proposed by Mazzone et al.~\cite{mazzone2024efficient} for CKKS.
We briefly describe their approach as we will later modify and optimize it for our use case.
Their core idea is to manipulate the encrypted input vector in such a way that all elements can be compared against each other with a single homomorphic evaluation of $\cmp$, exploiting the SIMD properties of the encryption scheme.
For instance, given a vector $\vector = (\vector_1, \vector_2, \vector_3)$, they produce
$$\vector_R = (\vector_1, \vector_2, \vector_3, \vector_1, \vector_2, \vector_3, \vector_1, \vector_2, \vector_3),$$
$$\vector_C = (\vector_1, \vector_1, \vector_1, \vector_2, \vector_2, \vector_2, \vector_3, \vector_3, \vector_3).$$
The comparison $\cmp(\vector_R, \vector_C)$ contains information about $\vector_i < \vector_j$ for all pairs $(\vector_i, \vector_j)$.
It is easier to visualize this by seeing $\vector_R, \vector_C$ as square matrices that have been encoded row-by-row into vectors:
$$
\vector_R =
\begin{pmatrix}
\vector_1 & \vector_2 & \vector_3 \\
\vector_1 & \vector_2 & \vector_3 \\
\vector_1 & \vector_2 & \vector_3 \\
\end{pmatrix},
\qquad
\vector_C =
\begin{pmatrix}
\vector_1 & \vector_1 & \vector_1 \\
\vector_2 & \vector_2 & \vector_2 \\
\vector_3 & \vector_3 & \vector_3 \\
\end{pmatrix}.
$$
The two encodings $\vector_R$ and $\vector_C$ are called the row and column encoding of $\vector$, respectively.
The following operations are available to work with an encrypted matrix $X$ in CKKS~\cite{mazzone2024efficient}:
\begin{itemize}
    \item $\maskR(X, i)$ extracts row $i$ by masking everything else, i.e., setting everything else to zero;
    \item $\sumR(X)$ sums all the rows together component-wise and stores the result in the first row;
    \item $\replR(X)$ assumes only the first row non-zero and replicates it by copying its values into the other rows;
    \item $\transR(X)$ assumes a square matrix with only the first row non-zero and transposes it (i.e., move it in the first column).
\end{itemize}
Similarly for the columns, we have $\maskC$, $\sumC$, $\replC$, $\transC$.
For these operations there are well-known algorithms in the literature that work recursively, and only require $\log(\vectorLength)$ rotations, where $\vectorLength$ is the number of rows/columns of the matrix~\cite{halevi2014algorithms, mazzone2024efficient}.
We collect their pseudocode in Appendix~\ref{app:rec_matrix_ops}.

Given an input vector $\vector$, Mazzone et al.'s approach for computing the argmin consists of three steps:
\begin{enumerate}
    \item \textbf{Encoding.} Compute the row and column encodings $\vector_R, \vector_C$.
    \item \textbf{Ranking.} Compare the two encodings to rank $\vector$.
    \item \textbf{Argmin.} Extract the index corresponding to rank $1$.
\end{enumerate}

\paragraph{Encoding}
Given an input vector $\vector$ encrypted as $\encVector$, we think of it as the first row of a null matrix.
The encoding $\vector_R$ is produced by simply applying \replR, while $\vector_C$ is produced by first transposing the initial vector to a column with \transR \, and then replicating it with \replC.

\paragraph{Ranking.}
Ranking associates the elements of a vector to their rank, that is the position they would have if the vector was sorted.
The component-wise comparison of $\vector_R > \vector_C$ is ideally a matrix with values in $\{0, 0.5, 1\}$, where each column $j$ contains information about the position of $\vector_j$ in the sorted array:
\begin{itemize}
    \item a number of ones equal to the number of elements smaller than $\vector_j$, and
    \item a 0.5 when compared against itself.
\end{itemize}
Thus, summing the elements in column $j$ and adding $0.5$ gives the rank of $\vector_j$ in the input vector.

\begin{algorithm}[t]
\caption{\textsf{Argmin}~\cite{mazzone2024efficient}}
\label{alg:argmin}
\begin{algorithmic}[1]
\Require $\encVector$ encryption of $\vector = (\vector_1, \dots, \vector_\vectorLength) \in \R^\vectorLength$.
\Ensure $\encArgminVar$ encryption of a vector in $\R^\vectorLength$ representing the one-hot encoding of the argmin of $\vector$.
\Statex \textbf{\textit{Encoding}}
\State $\encVector_R \gets \replR(\encVector)$
\State $\encVector_C \gets \replC(\transR(\encVector))$
\Statex \textbf{\textit{Ranking}}
\State $\encComparison \gets \cmp(\encVector_R, \encVector_C)$
\State $\encRanking \gets \sumR(\encComparison) + (0.5, \dots, 0.5)$
\Statex \textbf{\textit{Argmin}}
\State $\encArgminVar \gets \phi(\encRanking)$
\State \Return $\encArgminVar$
\end{algorithmic}
\end{algorithm}

\paragraph{Argmin.}
Now, the ranking $\ranking$ of $\vector$ will be a permutation of $(1, 2, \allowbreak\dots, \vectorLength)$, and we are interested in finding the position of rank $1$.
We are going to produce a one-hot encoding of this position by setting all non-zero values to $0$ and the only zero value to $1$.
To do so, Mazzone et al. suggest using a Chebyshev approximation of the indicator function around $1$.
However, this approach would be excessive in our case, since the vector length (i.e., the number of clusters) will be relatively low, typically on the order of tens.
Thus, we replace it with a simple equispaced nodes approximation
$$\phi(x) := \frac{1}{\prod_{j=2}^{\vectorLength}(1-j)} \prod_{j=2}^{\vectorLength}(x - j)$$
to reduce the number of homomorphic multiplications.
The pseudocode of the full algorithm is provided in Algorithm~\ref{alg:argmin}.

\paragraph{Handling Multiple Minima}
If two or more elements share the minimal value, the argmin algorithm returns a null vector.
This occurs because, in such cases, the ranking step maps all the minimal elements to the fractional rank $(u + 1) / 2$, where $u$ is the number of minimal elements.
As a consequence, the indicator function around 1 is not activated for any element.
For example, given the input vector $\vector = [10, 10, 30, 40]$, the resulting ranking is $\ranking = [1.5, 1.5, 3, 4]$, which is missing the rank 1.
The authors of~\cite{mazzone2024efficient} address this issue by introducing an offset vector that redistributes the fractional ranking of tied elements across the ranks they span.
However, we note that this adjustment is not required in our application.

In the clustering problem, having multiple minima implies that a data point is equidistant from two or more centroids.
In such cases, a null argmin would cause the point to be assigned to no cluster.
However, the probability of this scenario occurring with randomly initialized centroids is extremely low.
Our experimental evaluation confirms this observation.
Moreover, when a sufficient number of data points are present, the impact of a few unassigned points is negligible and does not affect the final outcome.


\section{K-Means Clustering}
\label{sec:main-construction}

We present the main construction of our protocol in the case of two parties --- Alice and Bob --- who want to partition a dataset of $\numPoints$ points in $\R^2$ into $\numClusters$ clusters.
We denote the points as $\point_i = (\point_i^A, \point_i^B)$ for $i = 1, \dots, \numPoints$.
Alice owns the first component $\point_i^A$ of each point, while Bob owns the second component $\point_i^B$.
We discuss how to extend our approach to an arbitrary number of dimensions in Section~\ref{sec:higher-dimensions}, and to an arbitrary number of parties in Section~\ref{sec:n-parties}.

Alice starts by initializing the centroids $\centroid_1, \dots, \centroid_\numClusters$.
Different initialization techniques are available for Lloyd's algorithm.
In this paper, we assume the data features are bounded in $[-\valueBound, \valueBound]$ for some $\valueBound > 0$, and Alice picks the initial set of centroids at random in $[-\valueBound, \valueBound]^2$, ensuring they are sufficiently spaced apart.
Notice that the centroids will be in plaintext the whole time, but protected by DP.
On the other side, Bob initializes the CKKS encryption scheme, by generating private, public, and evaluation keys.
Public and evaluation keys are then sent to Alice to enable her computing over Bob's encrypted data.
This step is data independent (offline phase, in MPC terminology) and can be done once and for all between the involved parties.

Bob encrypts his part of the dataset as a vector, splitting it across multiple ciphertexts if necessary.
To simplify the notation we indicate this as $\encPoint^B = \enc(\point_1^B, \dots, \point_\numPoints^B)$.
The encrypted $\encPoint^B$ is sent to Alice, who can now perform secure computations over Bob's data non-interactively.
Alice extracts each $\point_i^B$ from $\encPoint^B$ and encodes it in its own ciphertext by using masking and rotations: $\encPoint^B \cdot \delta_{j = i} \ll i$.
Then she uses $\replC$ to replicate the value of $\point_i^B$ $\numClusters$ times, obtaining $\encPoint_i^B$ as the encryption of $(\underbrace{\point_i^B, \dots, \point_i^B}_{\numClusters})$.

The following steps are then repeated:
\begin{enumerate}
    \item \textbf{Distance Computation.} For each point $i \in \{1, \dots, \numPoints\}$, Alice computes the distance between point $\point_i$ and each centroid $\centroid_j$ in one ciphertext.
    We employ the squared euclidean distance, since it is HE-friendly, that is:
    $$\distance_{ij} = (\point_i^A - \centroid_j^A)^2 + (\point_i^B - \centroid_j^B)^2 \enspace .$$
    To do so, Alice can compute Bob's part of the distance as $\encPoint_i^B - (\centroid_1^B, \dots, \centroid_\numClusters^B)$ and squaring the result.
    Then, she computes her part in plaintext, and adds it homomorphically to Bob's part.
    The resulting ciphertext $\encDistance_i$ encrypts the vector $(\distance_{i1}, \dots, \distance_{i\numClusters})$.
    \item \textbf{Cluster Computation.} Alice can now apply Algorithm~\ref{alg:argmin} to compute the argmin of the distances in each $\encDistance_i$.
    The result $\encArgminVar_i$ encrypts a one-hot encoding of $\bar{j}$, where $\centroid_{\bar{j}}$ is the closest centroid to $\point_i$.
    This concludes the centers-to-clusters phase, where each cluster $\cluster_j$ can be now defined as the set of points $\point_i$ for which $\centroid_j$ is the closest centroid.
    \item \textbf{Within-Cluster Mean.} By summing all the $\encArgminVar_i$ as $\encClusterSize = \sum_{i = 1}^{\numPoints}{\encArgminVar_i}$, Alice counts how many points belong to each cluster. In fact, $\encClusterSize$ is the encryption of $(|\cluster_1|, \dots, |\cluster_\numClusters|)$.
    She also computes the sum of the points' values in each cluster for both components as
    \begin{align*}
        \encSumPoints^A &= \sum_{i = 1}^{\numPoints}{\encArgminVar_i \cdot (\point_i^A, \dots, \point_i^A)} \\
        \encSumPoints^B &= \sum_{i = 1}^{\numPoints}{\encArgminVar_i \cdot \encPoint_i^B}
    \end{align*}
    which encrypt the vectors, respectively:
    \begin{align*}
        &\biggl(\,\sum_{\point_i \in \cluster_1}{\point_i^A}, \dots, \sum_{\point_i \in \cluster_\numClusters}{\point_i^A}\,\biggr) \\
        &\biggl(\,\sum_{\point_i \in \cluster_1}{\point_i^B}, \dots, \sum_{\point_i \in \cluster_\numClusters}{\point_i^B}\,\biggr) \enspace .
    \end{align*}
    \item \textbf{Noise Injection.} At this point, Alice could compute the new clusters as $\encSumPoints^A / \encClusterSize$ and $\encSumPoints^B / \encClusterSize$. However, this would require a very expensive division under encryption~\cite{cheon2019numerical}.
    Therefore, $\encSumPoints^A, \encSumPoints^B, \encClusterSize$ are sent to Bob, who can decrypt them and perform the division in plaintext.
    DP noise is added to these quantities by Alice before sending them, since they may leak information about the points $\point_i$:
    \begin{align*}
        \tilde{\encSumPoints}^A &= \encSumPoints^A + \gaussian(0, \sigma^2_\encSumPoints) \\
        \tilde{\encSumPoints}^B &= \encSumPoints^B + \gaussian(0, \sigma^2_\encSumPoints) \\
        \tilde{\encClusterSize} &= \encClusterSize + \gaussian(0, \sigma^2_\encClusterSize)
    \end{align*}
    The noise scales $\sigma_\encSumPoints, \sigma_\encClusterSize$ are discussed below.
    \item \textbf{Centroid Update.} Bob decrypts $\tilde{\encSumPoints}^A, \tilde{\encSumPoints}^B, \tilde{\encClusterSize}$ as $\tilde{\sumPoints}^A, \tilde{\sumPoints}^B, \tilde{\clusterSize}$ and performs the division
    $${\centroid'_j}^A = \tilde{\sumPoints}^A / \tilde{\clusterSize} \qquad {\centroid'_j}^B = \tilde{\sumPoints}^B / \tilde{\clusterSize}$$
    for $j \in \{1, \dots, \numClusters\}$.
    Finally, he sends the updated centroids back to Alice, who can start a new iteration of the algorithm.
\end{enumerate}

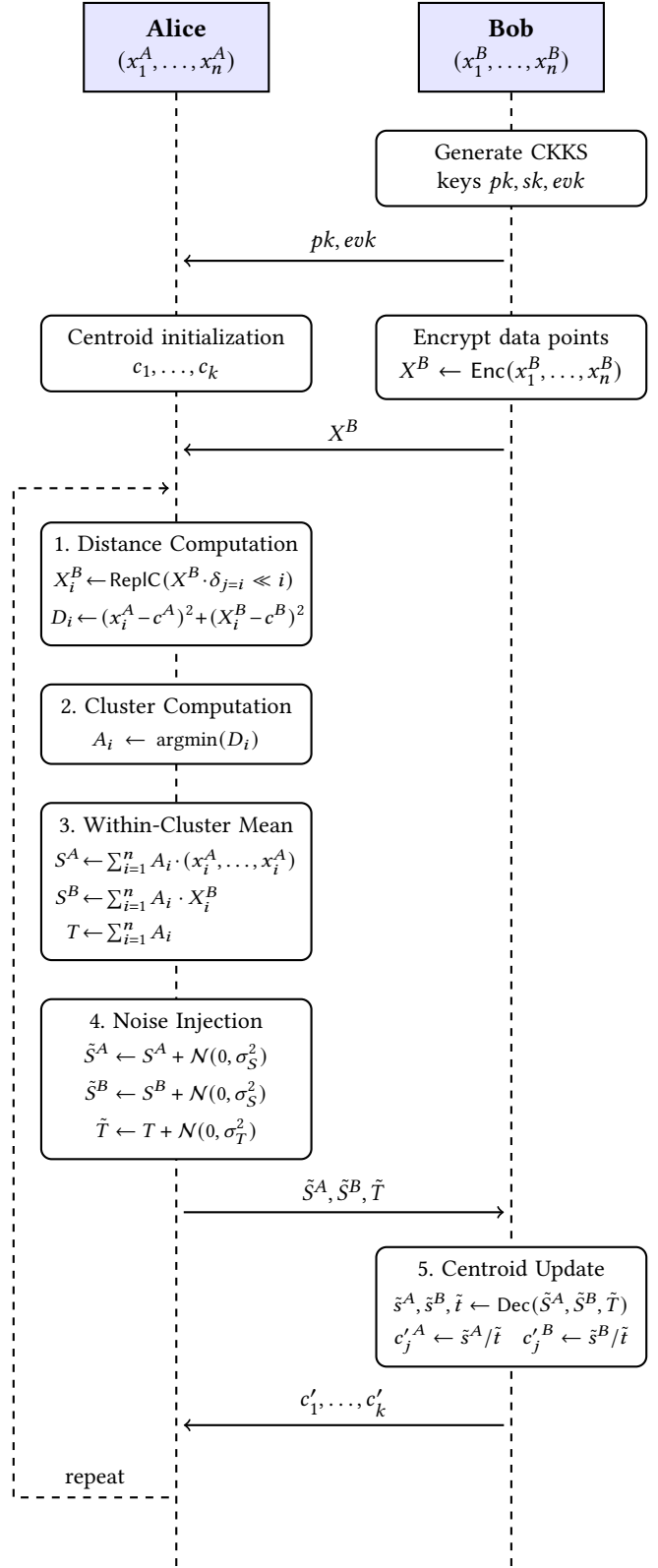
\begin{figure}
\centering
\begin{tikzpicture}[
    actor/.style={draw, thick, align=center, minimum width=2.5cm, minimum height=1.2cm, fill=blue!10},
    lifeline/.style={dashed, thick},
    msg/.style={->, thick, shorten >=0.1cm, shorten <=0.1cm},
    localstep/.style={align=center, text width=3.3cm, draw, thick, rounded corners, fill=white!10, inner sep=5pt}
]
\node[actor] (alice) {\large{\textbf{Alice}} \\[1pt] $(\point_1^A, \dots, \point_\numPoints^A)$};
\node[actor, right=2cm of alice] (bob) {\large{\textbf{Bob}} \\[1pt] $(\point_1^B, \dots, \point_\numPoints^B)$};
\draw[lifeline] (alice.south) -- ++(0,-20cm);
\draw[lifeline] (bob.south) -- ++(0,-20cm);
\node[below=0.5cm of bob.south, localstep, anchor=north] (bobSetup) {Generate CKKS keys $pk, sk, evk$};
\node[below=3cm of alice.south, localstep, anchor=north] (aliceCentroidInit) {Centroid initialization \\ $\centroid_1, \dots, \centroid_\numClusters$};
\node[below=3cm of bob.south, localstep, anchor=north] (bobEncrypt) {Encrypt data points \\[1.5pt] $\encPoint^B \gets \enc(\point_1^B, \dots, \point_\numPoints^B)$};
\node[below=5.8cm of alice.south, localstep, anchor=north] (aliceDist) {1. Distance Computation \\[-10pt] \small \begin{align*}
    &\encPoint_i^B \!\gets\! \replC(\encPoint^B \!\cdot\! \delta_{j = i} \ll i) \\
    &\!\encDistance_i \!\gets\! (\point_i^A \!-\! \centroid^A)^2 \!+\! (\encPoint_i^B \!-\! \centroid^B)^2
\end{align*}
};
\node[below=0.5cm of aliceDist.south, localstep, anchor=north] (aliceCluster) {2. Cluster Computation \\[1.5pt] \small $\encArgminVar_i \gets \argmin(\encDistance_i)$};
\node[below=0.5cm of aliceCluster.south, localstep, anchor=north] (aliceMean) {3. Within-Cluster Mean \\[-10pt] \small \begin{align*}
    \!\encSumPoints^A\! &\gets\! \textstyle\sum_{i = 1}^{\numPoints}{\encArgminVar_i \!\cdot\! (\point_i^A, \dots, \point_i^A)} \\
    \encSumPoints^B \!&\gets\! \textstyle\sum_{i = 1}^{\numPoints}{\encArgminVar_i \cdot \encPoint_i^B} \\
    \encClusterSize \!&\gets\! \textstyle\sum_{i = 1}^{\numPoints}{\encArgminVar_i}
    \end{align*}
};
\node[below=0.5cm of aliceMean.south, localstep, anchor=north] (aliceDP) {4. Noise Injection \\[-10pt] \small \begin{align*}
    \tilde{\encSumPoints}^A &\gets \encSumPoints^A + \gaussian(0, \sigma^2_\encSumPoints) \\
    \tilde{\encSumPoints}^B &\gets \encSumPoints^B + \gaussian(0, \sigma^2_\encSumPoints) \\
    \tilde{\encClusterSize} &\gets \encClusterSize + \gaussian(0, \sigma^2_\encClusterSize)
\end{align*}
};
\node[below=15.6cm of bob.south, localstep, anchor=north] (bobUpdate) {5. Centroid Update \\[3pt] \small $\tilde{\sumPoints}^A, \tilde{\sumPoints}^B, \tilde{\clusterSize} \gets \dec(\tilde{\encSumPoints}^A, \tilde{\encSumPoints}^B, \tilde{\encClusterSize})$ \\[1.5pt] ${\centroid'_j}^A \gets \tilde{\sumPoints}^A / \tilde{\clusterSize} \quad {\centroid'_j}^B \gets \tilde{\sumPoints}^B / \tilde{\clusterSize}$};
\draw[msg] ([yshift=-1.25cm]bob|-bobSetup) -- ([yshift=-1.25cm]alice|-bobSetup) node[midway, above] {$pk, evk$};
\draw[msg] ([yshift=-1.25cm]bob|-bobEncrypt) -- ([yshift=-1.25cm]alice|-bobEncrypt) node[midway, above] {$\encPoint^B$};
\draw[msg] ([yshift=-1.8cm]alice|-aliceDP) -- ([yshift=-1.8cm]bob|-aliceDP) node[midway, above] {$\tilde{\encSumPoints}^A, \tilde{\encSumPoints}^B, \tilde{\encClusterSize}$};
\draw[msg] ([yshift=-1.6cm]bob|-bobUpdate) -- ([yshift=-1.6cm]alice|-bobUpdate) node[midway, above] {$\centroid'_1, \dots, \centroid'_\numClusters$};
\draw[msg, dashed] ([yshift=-19cm]alice.south) -| ++(-2.2cm,0) node[near start, above] {repeat} |- ([yshift=-5.35cm]alice.south);
\end{tikzpicture}
\caption{Secure k-means clustering protocol steps.}
\label{fig:main-construction}
\end{figure}

These steps are repeated until some convergence condition is satisfied, for example until the centroids are sufficiently stable, or a fixed number of rounds is executed.
Besides the CKKS keys, the communication involved in our solution consists of the encrypted points $\encPoint^B$ sent from Bob at the beginning, and of the centroids sent back and forth at the end of each iteration.
Hence, the communication complexity of our approach is $O(\numPoints + \numClusters \numRounds)$ where $\numRounds$ is the total number of iterations.
On the other hand, the computation complexity is the same as the plaintext Lloyd's algorithm, that is $O(\numPoints \numClusters \numRounds)$, but obviously with higher multiplicative constants.
Figure~\ref{fig:main-construction} shows a schematic overview of the protocol.

\paragraph{Differential Privacy}
To calculate the noise scale in Step 4, we must first determine the sensitivity of the quantities involved.
Given two datasets that differ by at most one record, the size of any cluster can vary by at most one, as that record may switch from one cluster to another.
Thus, the sensitivity of $\clusterSize$ is $1$.
Similarly, the sensitivity of the cluster sums $\sumPoints^A$ and $\sumPoints^B$ is $2\valueBound$.
Given a privacy budget $\epsilon$ and a failure probability $\delta$, we distribute these parameters evenly across the rounds, using either the simple or advanced composition method, depending on which yields the larger privacy budget.
Then, Alice adds noise following the Gaussian mechanism described in Section~\ref{sec:background:dp}, using
\begin{align*}
    \sigma_\encSumPoints &= \sqrt{2 \log(1.25 / \delta')} \, / \epsilon' \\
    \sigma_\encClusterSize &= \sqrt{2 \log(1.25 / \delta')} \, 2 \valueBound / \epsilon'
\end{align*}
where $(\epsilon', \delta')$ are the per-round privacy parameters.

\subsection{Optimizations}

While the presented protocol is efficient in terms of communication, its computational overhead in the current form is too high for practical deployment.
Specifically, the argmin computation requires several seconds per data point, making the cost prohibitive for large datasets.
We propose multiple optimizations to reduce the computation time, enabling the protocol to scale to practical use cases.

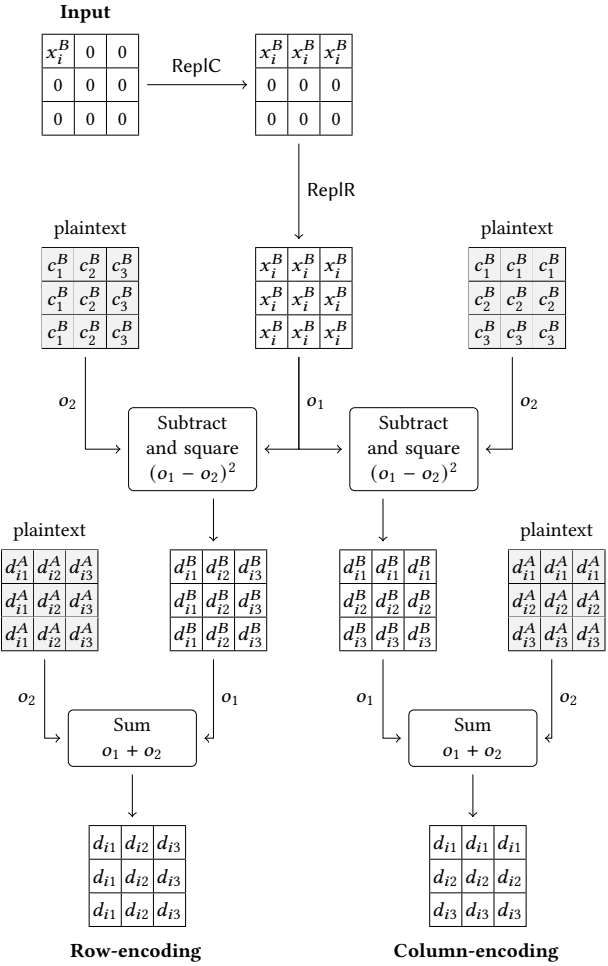
\begin{figure}
    \centering
    \scalebox{0.95}{
    \begin{tikzpicture}[node distance=85pt]

    \tikzstyle{every node}=[font=\small]

    \node (table1) {
    \renewcommand{\tabcolsep}{\tabcolsepValue}
    \renewcommand{\arraystretch}{\arraystretchValue}
    \begin{tabular}{|P{10pt}|P{10pt}|P{10pt}|}
        \hline
        $\point_i^B$ & 0 & 0 \\ \hline
        0 & 0 & 0 \\ \hline
        0 & 0 & 0 \\ \hline
    \end{tabular}
    };
    
    \node[right of=table1] (table2) {
    \renewcommand{\tabcolsep}{\tabcolsepValue}
    \renewcommand{\arraystretch}{\arraystretchValue}
    \begin{tabular}{|P{10pt}|P{10pt}|P{10pt}|}
        \hline
        $\point_i^B$ & $\point_i^B$ & $\point_i^B$ \\ \hline
        0 & 0 & 0 \\ \hline
        0 & 0 & 0 \\ \hline
    \end{tabular}
    };

    \node[below of=table2] (table3) {
    \renewcommand{\tabcolsep}{\tabcolsepValue}
    \renewcommand{\arraystretch}{\arraystretchValue}
    \begin{tabular}{|P{10pt}|P{10pt}|P{10pt}|}
        \hline
        $\point_i^B$ & $\point_i^B$ & $\point_i^B$ \\ \hline
        $\point_i^B$ & $\point_i^B$ & $\point_i^B$ \\ \hline
        $\point_i^B$ & $\point_i^B$ & $\point_i^B$ \\ \hline
    \end{tabular}
    };

    \node[left of=table3] (table4) {
    \renewcommand{\tabcolsep}{\tabcolsepValue}
    \renewcommand{\arraystretch}{\arraystretchValue}
    \begin{tabular}{|P{10pt}|P{10pt}|P{10pt}|}
        \hline
        \cellcolor{gray!10}$\centroid_1^B$ & \cellcolor{gray!10}$\centroid_2^B$ & \cellcolor{gray!10}$\centroid_3^B$ \\ \hline
        \cellcolor{gray!10}$\centroid_1^B$ & \cellcolor{gray!10}$\centroid_2^B$ & \cellcolor{gray!10}$\centroid_3^B$ \\ \hline
        \cellcolor{gray!10}$\centroid_1^B$ & \cellcolor{gray!10}$\centroid_2^B$ & \cellcolor{gray!10}$\centroid_3^B$ \\ \hline
    \end{tabular}
    };

    \node[right of=table3] (table5) {
    \renewcommand{\tabcolsep}{\tabcolsepValue}
    \renewcommand{\arraystretch}{\arraystretchValue}
    \begin{tabular}{|P{10pt}|P{10pt}|P{10pt}|}
        \hline
        \cellcolor{gray!10}$\centroid_1^B$ & \cellcolor{gray!10}$\centroid_1^B$ & \cellcolor{gray!10}$\centroid_1^B$ \\ \hline
        \cellcolor{gray!10}$\centroid_2^B$ & \cellcolor{gray!10}$\centroid_2^B$ & \cellcolor{gray!10}$\centroid_2^B$ \\ \hline
        \cellcolor{gray!10}$\centroid_3^B$ & \cellcolor{gray!10}$\centroid_3^B$ & \cellcolor{gray!10}$\centroid_3^B$ \\ \hline
    \end{tabular}
    };
    
    \node[draw, rounded corners=2pt, below left=20pt and -8pt of table3, text width=45pt, align=center] (minus1) {
    Subtract and square $(o_1 - o_2)^2$
    };

    \node[draw, rounded corners=2pt, below right=20pt and -4.5pt of table3, text width=45pt, align=center] (minus2) {
    Subtract and square $(o_1 - o_2)^2$
    };

    \node[below right=20pt and -42pt of minus1] (table6) {
    \renewcommand{\tabcolsep}{\tabcolsepValue}
    \renewcommand{\arraystretch}{\arraystretchValue}
    \begin{tabular}{|P{10pt}|P{10pt}|P{10pt}|}
        \hline
        $\distance_{i1}^B$ & $\distance_{i2}^B$ & $\distance_{i3}^B$ \\ \hline
        $\distance_{i1}^B$ & $\distance_{i2}^B$ & $\distance_{i3}^B$ \\ \hline
        $\distance_{i1}^B$ & $\distance_{i2}^B$ & $\distance_{i3}^B$ \\ \hline
    \end{tabular}
    };

    \node[below left=20pt and -38pt of minus2] (table7) {
    \renewcommand{\tabcolsep}{\tabcolsepValue}
    \renewcommand{\arraystretch}{\arraystretchValue}
    \begin{tabular}{|P{10pt}|P{10pt}|P{10pt}|}
        \hline
        $\distance_{i1}^B$ & $\distance_{i1}^B$ & $\distance_{i1}^B$ \\ \hline
        $\distance_{i2}^B$ & $\distance_{i2}^B$ & $\distance_{i2}^B$ \\ \hline
        $\distance_{i3}^B$ & $\distance_{i3}^B$ & $\distance_{i3}^B$ \\ \hline
    \end{tabular}
    };

    \node[left=18pt of table6] (table8) {
    \renewcommand{\tabcolsep}{\tabcolsepValue}
    \renewcommand{\arraystretch}{\arraystretchValue}
    \begin{tabular}{|P{10pt}|P{10pt}|P{10pt}|}
        \hline
        \cellcolor{gray!10}$\distance_{i1}^A$ & \cellcolor{gray!10}$\distance_{i2}^A$ & \cellcolor{gray!10}$\distance_{i3}^A$ \\ \hline
        \cellcolor{gray!10}$\distance_{i1}^A$ & \cellcolor{gray!10}$\distance_{i2}^A$ & \cellcolor{gray!10}$\distance_{i3}^A$ \\ \hline
        \cellcolor{gray!10}$\distance_{i1}^A$ & \cellcolor{gray!10}$\distance_{i2}^A$ & \cellcolor{gray!10}$\distance_{i3}^A$ \\ \hline
    \end{tabular}
    };

    \node[right=18pt of table7] (table9) {
    \renewcommand{\tabcolsep}{\tabcolsepValue}
    \renewcommand{\arraystretch}{\arraystretchValue}
    \begin{tabular}{|P{10pt}|P{10pt}|P{10pt}|}
        \hline
        \cellcolor{gray!10}$\distance_{i1}^A$ & \cellcolor{gray!10}$\distance_{i1}^A$ & \cellcolor{gray!10}$\distance_{i1}^A$ \\ \hline
        \cellcolor{gray!10}$\distance_{i2}^A$ & \cellcolor{gray!10}$\distance_{i2}^A$ & \cellcolor{gray!10}$\distance_{i2}^A$ \\ \hline
        \cellcolor{gray!10}$\distance_{i3}^A$ & \cellcolor{gray!10}$\distance_{i3}^A$ & \cellcolor{gray!10}$\distance_{i3}^A$ \\ \hline
    \end{tabular}
    };

    \node[draw, rounded corners=2pt, below left=20pt and -18pt of table6, text width=45pt, align=center] (sum1) {
    Sum \\ $o_1 + o_2$
    };

    \node[draw, rounded corners=2pt, below right=20pt and -14.5pt of table7, text width=45pt, align=center] (sum2) {
    Sum \\ $o_1 + o_2$
    };

    \node[below=20pt of sum1] (table10) {
    \renewcommand{\tabcolsep}{\tabcolsepValue}
    \renewcommand{\arraystretch}{\arraystretchValue}
    \begin{tabular}{|P{10pt}|P{10pt}|P{10pt}|}
        \hline
        $\distance_{i1}$ & $\distance_{i2}$ & $\distance_{i3}$ \\ \hline
        $\distance_{i1}$ & $\distance_{i2}$ & $\distance_{i3}$ \\ \hline
        $\distance_{i1}$ & $\distance_{i2}$ & $\distance_{i3}$ \\ \hline
    \end{tabular}
    };

    \node[below=20pt of sum2] (table11) {
    \renewcommand{\tabcolsep}{\tabcolsepValue}
    \renewcommand{\arraystretch}{\arraystretchValue}
    \begin{tabular}{|P{10pt}|P{10pt}|P{10pt}|}
        \hline
        $\distance_{i1}$ & $\distance_{i1}$ & $\distance_{i1}$ \\ \hline
        $\distance_{i2}$ & $\distance_{i2}$ & $\distance_{i2}$ \\ \hline
        $\distance_{i3}$ & $\distance_{i3}$ & $\distance_{i3}$ \\ \hline
    \end{tabular}
    };

    \node[above=-2pt of table1] {\textbf{Input}};
    \node[above=-3pt of table4] {\,\,\,plaintext};
    \node[above=-3pt of table5] {\,\,\,plaintext};
    \node[above=-3pt of table8] {\,\,\,plaintext};
    \node[above=-3pt of table9] {\,\,\,plaintext};
    \node[below=0pt of table10] {\textbf{\,\,Row-encoding}};
    \node[below=0pt of table11] {\textbf{\,\,Column-encoding}};

    \draw[->, shorten >=-4pt] (table1) -- (table2) node[pos=0.55, above] {\replC};
    \draw[->] (table2) -- (table3) node[midway, right] {\replR};
    \draw[->, shorten >=3pt] (table3) |- (minus1) node[near start, right] {$o_1$};
    \draw[->, shorten >=3pt] (table3) |- (minus2);
    \draw[->, shorten >=3pt] (table4) |- (minus1) node[near start, left] {$o_2$};
    \draw[->, shorten >=3pt] (table5) |- (minus2) node[near start, right] {$o_2$};
    \draw[->, shorten <=3pt] (minus1.south -| table6.north) -- (table6.north);
    \draw[->, shorten <=3pt] (minus2.south -| table7.north) -- (table7.north);
    \draw[->, shorten >=3pt] (table6) |- (sum1) node[near start, right] {$o_1$};
    \draw[->, shorten >=3pt] (table7) |- (sum2) node[near start, left] {$o_1$};
    \draw[->, shorten >=3pt] (table8) |- (sum1) node[near start, left] {$o_2$};
    \draw[->, shorten >=3pt] (table9) |- (sum2) node[near start, right] {$o_2$};
    \draw[->, shorten <=3pt] (sum1.south -| table10.north) -- (table10.north);
    \draw[->, shorten <=3pt] (sum2.south -| table11.north) -- (table11.north);

    \end{tikzpicture}
    }
    
    \caption{Schematic example of optimized encoding for $\numClusters = 3$.}
    \label{fig:optimizing-argmin-encoding}
\end{figure}

\paragraph{Optimizing the Argmin Encoding Phase}
First, we notice that the part of the distance $\distance_{ij} = (\point_i^A - \centroid_j^A)^2 + (\point_i^B - \centroid_j^B)^2$ that varies across rounds consists only of the centroids $\centroid_j^A, \centroid_j^B$, which are in plaintext.
Thus, instead of first computing $\encDistance_i$ as encrypted vector and then replicating and transposing it during the argmin encoding phase, we can take a more efficient approach that we present in Figure~\ref{fig:optimizing-argmin-encoding}.
Using $\replR$, we replicate $\encPoint_i^B$ into a square matrix containing only the value $\point_i^B$.
Then, we subtract a replicated encoding of the centroids $\centroid^B$, first using row encoding and then column encoding, both in plaintext.
The result is squared to get Bob's part of the distance $\distance^B = (\point_i^B - \centroid_j^B)^2$, and added to Alice's part of the distance $\distance^A = (\point_i^A - \centroid_j^A)^2$, which was computed in plaintext and properly encoded.
This process produces the necessary row and column encodings required for the comparison step in the argmin computation.
Moreover, since Bob’s component of the data points $\point_i^B$ remains constant across all rounds, we can perform the replication step just once at the start.
As a result, the argmin encoding phase is reduced to a straightforward plaintext encoding operation, which is orders-of-magnitude faster than its counterpart under encryption.

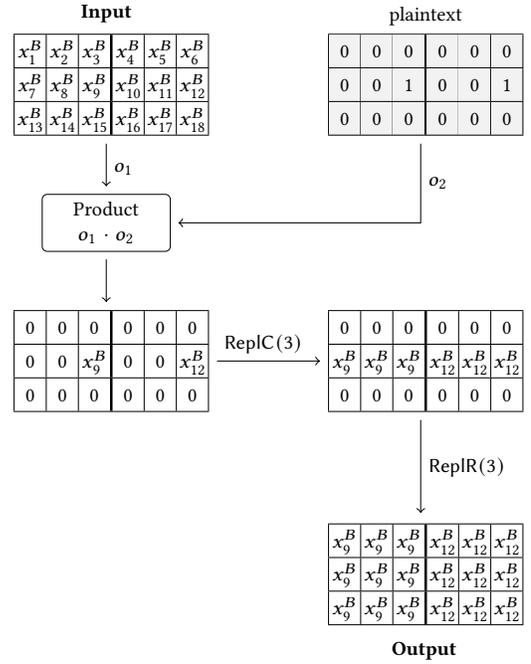
\begin{figure}
    \centering
    \scalebox{0.95}{
    \begin{tikzpicture}[node distance=85pt]

    \tikzstyle{every node}=[font=\small]

    \node (table1) {
    \renewcommand{\tabcolsep}{\tabcolsepValue}
    \renewcommand{\arraystretch}{\arraystretchValue}
    \begin{tabular}{|P{10pt}|P{10pt}|P{10pt}"P{10pt}|P{10pt}|P{10pt}|}
        \hline
        $\point_1^B$ & $\point_2^B$ & $\point_3^B$ & $\point_4^B$ & $\point_5^B$ & $\point_6^B$ \\ \hline
        $\point_7^B$ & $\point_8^B$ & $\point_9^B$ & $\point_{10}^B$ & $\point_{11}^B$ & $\point_{12}^B$ \\ \hline
        $\point_{13}^B$ & $\point_{14}^B$ & $\point_{15}^B$ & $\point_{16}^B$ & $\point_{17}^B$ & $\point_{18}^B$ \\ \hline
    \end{tabular}
    };
    
    \node[right=37pt of table1] (table2) {
    \renewcommand{\tabcolsep}{\tabcolsepValue}
    \renewcommand{\arraystretch}{\arraystretchValue}
    \begin{tabular}{|P{10pt}|P{10pt}|P{10pt}"P{10pt}|P{10pt}|P{10pt}|}
        \hline
        \cellcolor{gray!10}0 & \cellcolor{gray!10}0 & \cellcolor{gray!10}0 & \cellcolor{gray!10}0 & \cellcolor{gray!10}0 & \cellcolor{gray!10}0 \\ \hline
        \cellcolor{gray!10}0 & \cellcolor{gray!10}0 & \cellcolor{gray!10}1 & \cellcolor{gray!10}0 & \cellcolor{gray!10}0 & \cellcolor{gray!10}1 \\ \hline
        \cellcolor{gray!10}0 & \cellcolor{gray!10}0 & \cellcolor{gray!10}0 & \cellcolor{gray!10}0 & \cellcolor{gray!10}0 & \cellcolor{gray!10}0 \\ \hline
    \end{tabular}
    };

    \node[draw, rounded corners=2pt, below=20pt of table1, text width=45pt, align=center] (product) {
    Product $o_1 \cdot o_2$
    };

    \node[below=20pt of product] (table3) {
    \renewcommand{\tabcolsep}{\tabcolsepValue}
    \renewcommand{\arraystretch}{\arraystretchValue}
    \begin{tabular}{|P{10pt}|P{10pt}|P{10pt}"P{10pt}|P{10pt}|P{10pt}|}
        \hline
        0 & 0 & 0 & 0 & 0 & 0 \\ \hline
        0 & 0 & $\point_9^B$ & 0 & 0 & $\point_{12}^B$ \\ \hline
        0 & 0 & 0 & 0 & 0 & 0 \\ \hline
    \end{tabular}
    };

    \node[right=37pt of table3] (table4) {
    \renewcommand{\tabcolsep}{\tabcolsepValue}
    \renewcommand{\arraystretch}{\arraystretchValue}
    \begin{tabular}{|P{10pt}|P{10pt}|P{10pt}"P{10pt}|P{10pt}|P{10pt}|}
        \hline
        0 & 0 & 0 & 0 & 0 & 0 \\ \hline
        $\point_9^B$ & $\point_9^B$ & $\point_9^B$ & $\point_{12}^B$ & $\point_{12}^B$ & $\point_{12}^B$ \\ \hline
        0 & 0 & 0 & 0 & 0 & 0 \\ \hline
    \end{tabular}
    };

    \node[below of=table4] (table5) {
    \renewcommand{\tabcolsep}{\tabcolsepValue}
    \renewcommand{\arraystretch}{\arraystretchValue}
    \begin{tabular}{|P{10pt}|P{10pt}|P{10pt}"P{10pt}|P{10pt}|P{10pt}|}
        \hline
        $\point_9^B$ & $\point_9^B$ & $\point_9^B$ & $\point_{12}^B$ & $\point_{12}^B$ & $\point_{12}^B$ \\ \hline
        $\point_9^B$ & $\point_9^B$ & $\point_9^B$ & $\point_{12}^B$ & $\point_{12}^B$ & $\point_{12}^B$ \\ \hline
        $\point_9^B$ & $\point_9^B$ & $\point_9^B$ & $\point_{12}^B$ & $\point_{12}^B$ & $\point_{12}^B$ \\ \hline
    \end{tabular}
    };

    \node[above=-2pt of table1] {\textbf{Input}};
    \node[above=-3pt of table2] {\,\,\,plaintext};
    \node[below=0pt of table5] {\textbf{\,\,Output}};

    \draw[->, shorten >=3pt] (table1) -- (product) node[midway, right] {$o_1$};
    \draw[->, shorten >=3pt] (table2) |- (product) node[near start, right] {$o_2$};
    \draw[->, shorten <=3pt] (product.south -| table3.north) -- (table3.north);
    \draw[->, shorten >=-4pt] (table3) -- (table4) node[midway, above] {$\replC(3)$};
    \draw[->, shorten >=1pt] (table4) -- (table5) node[midway, right] {$\replR(3)$};

    \end{tikzpicture}
    }
    
    \caption{Schematic example of extracting and re-encoding multiple points from $\encPoint^B$ in one ciphertext. In this example we have $\numClusters = 3$ clusters, $18$ ciphertext slots, and we extract and re-encode the $6$th element of each $\numClusters \times \numClusters$ square.}
    \label{fig:multiple-argmin}
\end{figure}

\paragraph{Processing Multiple Argmin in One Ciphertext}
Another optimization involves making full use of all available slots in a ciphertext to parallelize multiple argmin computations.
The key observation is that homomorphic operations take the same amount of time regardless of how many ciphertext slots are actually being used.
In our implementation, we use a CKKS ring dimension of $\ringDim = 2^{15}$, which allows each ciphertext to encode up to $2^{14}$ elements.
However, as described so far, we are only storing $\bar{\numClusters}^2$ elements per ciphertext, where $\bar{\numClusters}$ is the least power of two greater or equal than $\numClusters$.
For typical use cases, where $\numClusters$ is usually less than 10, this results in many unused slots.
For example, for $\numClusters = 3$, we would only use 16 slots out of 16,384, leaving \mytilde 99.98\% of them unused.

To exploit this unused space, we pack multiple points $\point_i^B$ into the same ciphertext and process them concurrently.
Specifically, we can pack up to $\floor{2^{\ringDim / 2} / \bar{\numClusters}^2}$ points in one ciphertext, as each point requires a $\bar{\numClusters} \times \bar{\numClusters}$ matrix for computing the argmin.
These matrices are encoded row-wise and stacked horizontally within the ciphertext, meaning the ciphertext first stores the first row of each matrix, then the second row, and so on.
Matrix operations such as $\replR, \replC, \sumR, \sumC$ are easily adapted to work in this setting.
Figure~\ref{fig:multiple-argmin} shows how to extract and re-encode multiple elements (two in the example) from Bob's compact encoding $\encPoint^B$.
In this example, we assume the ciphertext has only 18 slots for simplicity. Given 18 elements from Bob, we generate 9 ciphertexts, each encoding 2 elements.
Additionally, this example assumes that no padding to a power of two is needed, which turns out to be true, as we discuss below.

After the argmin is computed, each ciphertext will encrypt the closest centroid of multiple points, as a concatenation of one-hot encoding vectors.
Step 3 of our solution (within-cluster means) is adapted accordingly by performing both a sum over ciphertexts and a sum within ciphertexts with $\sumC$.

\paragraph{Eliminating the Need for Padding}
We can further optimize the use of ciphertext slots by removing the need for padding.
Recursive operations on encrypted matrices are efficient, using only a logarithmic number of homomorphic additions and rotations.
However, they require the matrix to be padded to the nearest power of two, which wastes ciphertext slots.
To address this, we modify the replication and summation algorithms to handle the case of $\numClusters$ not being a power of two.
For the replication algorithm, we limit the recursion to the largest power of two smaller than $\numClusters$, then we fill the remaining slots using the intermediate results of the recursion.
For example, to replicate a value 14 times, we first double it recursively up to 8 slots (computing replications for 2, 4, and 8 slots) and then add the replications for 2 and 4 slots to reach the full 14 slots.
This way we only use $14^2 = 196$ slots per point, instead of $16^2 = 256$, saving around 23\% slots.
Additionally, we can make the replication start from any initial position of the non-zero element, avoiding an extra rotation to bring it to position zero.
We apply a similar method for summation.
In the worst case, this approach requires $2 \floor{\log(\numClusters)} - 1$ rotations, which is more than the $\ceil{\log(\numClusters)}$ rotations required with padding.
However, our experiments show that this increased cost is outweighed by the benefits of saving slots, which allows more points to be processed in parallel.
The pseudocode for this approach is provided in Appendix~\ref{app:repl-no-padding} as Algorithm~\ref{alg:super-repl}.

\paragraph{Special Case: $\numClusters = 2$}
When working with only two clusters, the process can be further optimized by performing the argmin computation directly on the compact encoding.
In this scenario, only a single comparison is required for the argmin, making the use of the argmin approach from Mazzone et al.~\cite{mazzone2024efficient} pointless.
Instead, we compare the following quantities directly, without any re-encoding:
\begin{align*}
    &\big(\point^A - (\centroid_1^A, \dots, \centroid_1^A)\big)^2 + \big(\encPoint^B - (\centroid_1^B, \dots, \centroid_1^B)\big)^2 \enspace \text{and} \\
    &\big(\point^A - (\centroid_2^A, \dots, \centroid_2^A)\big)^2 + \big(\encPoint^B - (\centroid_2^B, \dots, \centroid_2^B)\big)^2 \enspace .
\end{align*}
The output is a bitmask $\encArgminVar$ indicating which points are closer to $\centroid_2$.
Conversely, $1 - \encArgminVar$ indicates which points are closer to $\centroid_1$.
This approach allows processing four times as many points in parallel and eliminates the need for the $\sumR$ and $\phi$ operations in the argmin computation.
The rest of the algorithm proceeds as usual.

\subsection{Extension to Higher Dimensions}
\label{sec:higher-dimensions}
Extending our approach to arbitrary dimensions is straightforward and introduces only minor overhead.
Instead of two-dimensional points, we now consider points $\point_i \in \R^\numDims$ for an arbitrary $\numDims \ge 2$.
Let $\numDims_A$ and $\numDims_B$ be the number of features owned by Alice and Bob, respectively, such that $\numDims = \numDims_A + \numDims_B$.
The primary adjustment to our approach is that the distance computation now involves $\numDims_A$ plaintext features and $\numDims_B$ encrypted features.
$$\encDistance_i \gets \sum_{l = 1}^{\numDims^A}{\big(\point_i^l - \centroid^l\big)^2} + \sum_{l = 1}^{\numDims^B}{\big(\encPoint_i^l - \centroid^l\big)^2}$$
Additionally, within-cluster means must be computed for $\numDims$ dimensions instead of 2.

The computational overhead is negligible, as the most expensive operation --- the argmin computation --- is agnostic to the number of data features.
The communication cost naturally increases with the number of dimensions, as each point must now be represented by $\numDims$ values.
Specifically:
\begin{itemize}
    \item The size of Bob's encrypted data scales linearly with the number of features he owns, namely $\numDims_B$.
    \item The size of the centroids scales linearly with the total number of features, namely $\numDims$.
\end{itemize}
This results in an overall communication complexity of $O(\numPoints \numDims_B + \numClusters \numRounds \numDims)$.
As a consequence, to minimize the communication cost, it is preferable for the party with more features to play the role of Alice.


\section{Generalization to \numParties \ Parties}
\label{sec:n-parties}

Our solution can be extended to support an arbitrary number of parties.
We outline two models for this extension:
\begin{itemize}
    \item \textbf{server-aided model}: computations are outsourced to two non-colluding parties, taking on the roles of Alice and Bob;
    \item \textbf{multiparty computation model}: all parties participate directly in the protocol without requiring trust that two of them will not collude.
\end{itemize}

\subsection{Server-Aided Model}

In the server-aided model, one party is designated as the computing party, acting as Alice, while another party takes on the role of Bob.
At the beginning of the protocol, Bob generates the HE keys and broadcasts the public key to all other parties, enabling them to encrypt their data and send it to Alice.
Alice and Bob then execute the protocol as previously described just among them two.
After the final round, the resulting clusters are shared with all parties.

This model requires a non-collusion assumption between Alice and Bob, as Alice holds the encrypted data from all parties, while Bob has the ability to decrypt it.
From a communication and computation cost perspective, this approach is equivalent to a two-party scenario where Bob owns all the features from the other parties.
The only overhead is the cost associated with broadcasting the public key and the final clusters to all participants.

\subsection{Multiparty Computation Model}

In the multiparty computation model, one party is designated as the computing party and takes on the role of Alice, while the role of Bob is distributed across all the other parties.
The key idea is that no single party possesses the secret key for the HE scheme, meaning that even if a party colludes with the computing server, they cannot decrypt the data of the other parties.
This can be achieved using multiparty homomorphic encryption (MHE)~\cite{mouchet2021multiparty}.

With MHE, each non-computing party holds only a share of the secret key.
As a result, while data encryption and homomorphic operations can be performed by anyone, decryption requires collaboration among all parties.
Specifically, after the setup phase, each party encrypts its data and sends it to Alice independently.
When decrypting intermediate cluster means and sizes, Alice broadcasts the relevant ciphertexts to all parties and receives decryption shares from each of them.
By aggregating these shares, Alice obtains the plaintexts needed to update the centroids after performing the necessary divisions.

With this model, the communication size scales linearly with the number of parties, as each must contribute to the decryption.
However, the number of rounds remains the same since all communication can occur in parallel.
Computationally, the overhead compared to the two-party setting is negligible, with only a slight increase due to the aggregation of decryption shares.

Since the secret key is shared additively among the parties, successful decryption requires the collaboration of all participants.
This ensures that even if all but one of the non-computing parties collude with Alice, they cannot access the data of the remaining honest party.

For scenarios where some non-computing parties prefer not to participate in every round and wish to simply outsource their data and receive results at the end, a threshold secret sharing scheme can be used.
This approach relaxes the strict $\numParties$-out-of-$\numParties$ non-collusion property, ensuring that decryption requires only a subset of the non-computing parties to be online.


\begin{table*}
  \caption{Network configurations.}
  \label{tab:network_configs}
  \begin{tabular}{llrrrrrrr}
    \toprule
    Identifier
    & Description
    & Bandwidth
    & Delay
    & Jitter
    & Packet Loss
    & Burst Size
    & Quantum
    & r2q
    \\
    
    & 
    & (Mbps)
    & (ms)
    & (ms)
    & (\%)
    & (KB)
    & (bytes)
    & 
    \\
    \midrule
    LAN500      & LAN (low)                       & 500   & 1     & 0.2   & -     & -     & 1500  & 10    \\
    LAN1000     & LAN (medium)                    & 1000  & 0.3   & 0.02  & -     & -     & 3000  & 15    \\
    LAN10000    & LAN (high)                      & 10000 & 0.1   & 0.01  & -     & -     & 9000  & 25    \\
    regWAN100   & Regional WAN (low)              & 100   & 20    & 15    & 0.1   & 500   & 1200  & 10    \\
    regWAN250   & Regional WAN (medium)           & 250   & 15    & 5     & 0.1   & 1000  & 1500  & 20   \\
    regWAN500   & Regional WAN (high)             & 500   & 10    & 2     & 0.1   & 1500  & 1500  & 25   \\
    ccWAN50     & Cross-continental WAN (low)     & 50    & 150   & 25    & 0.5   & 500   & 1000  & 10   \\
    ccWAN100    & Cross-continental WAN (medium)  & 100   & 120   & 15    & 0.3   & 1000  & 1000  & 15   \\
    ccWAN200    & Cross-continental WAN (high)    & 200   & 100   & 10    & 0.2   & 2000  & 1200  & 20   \\
    crpWAN500   & Corporate WAN (high resilience) & 500   & 50    & 5     & 0.1   & 1000  & 1500  & 15   \\
  \bottomrule
\end{tabular}
\end{table*}

\section{Experimental Results}

We implement and evaluate our approach on real-world and synthetic datasets to demonstrate its scalability and performance under different network configurations.
Our solution is compared against state-of-the-art approaches for MPC~\cite{mohassel2020practical} and DP~\cite{li2022differentially} in terms of runtime and accuracy.

\subsection{Experimental Setup}

Our solution is built on top of the CKKS implementation provided by OpenFHE~\cite{openFHE}\footnote{\url{https://github.com/openfheorg/openfhe-development}}.
We employ a scaling factor of 32 bits, while the ring dimension is set to $2^{15}$, hence each ciphertext can encode a vector of up to $2^{14}$ elements.
The multiplicative depth ranges from 13 for 2 clusters to 18 for 15 clusters.
All parameters are chosen in accordance with the Homomorphic Encryption Standard to ensure 128-bit security~\cite{albrecht2015concrete, HomomorphicEncryptionSecurityStandard}.
We make our code available open-source at \url{\repository}.

Experiments are conducted on a machine equipped with 8x Intel Xeon Platinum 8276 processors and 6 TB of RAM.
We run all parties as distinct processes on the same machine, using the \texttt{tc} (traffic control) command to configure network settings such as bandwidth, delay, and packet loss.
To simulate a variety of real-world scenarios, we test ten different network configurations, ranging from a slow 20 Mbps WAN connection to a high-speed 10 Gbps LAN connection.
These configurations are detailed in Table~\ref{tab:network_configs}.

\subsection{Datasets}

We describe the datasets used for our experimental evaluation.

\paragraph{Loan}
This consists of 60,000 records from a dataset released by Home Credit.\footnote{\url{https://www.kaggle.com/competitions/home-credit-default-risk}}
The features used are 16 including credit amount, family size, housing characteristics, and social circle statistics.

\paragraph{Taxi}
This consists of 100,000 records from a dataset released by the NYC Taxi and Limousine Commission in 2016.\footnote{\url{https://www.nyc.gov/site/tlc/about/tlc-trip-record-data.page}}
The features used are 8 including pickup time, geo-coordinates, and number of passengers

\vspace{5pt}

These two datasets are also used in~\cite{li2022differentially}, with a number of clusters equal to 5 for both.
For consistency reasons, we pre-process them as indicated by the authors, namely by turning non-numerical features into numerical, clipping or discarding outliers over the 95th percentile, and normalizing each feature in $[0, 1]$.

\paragraph{Bank}
This consists of just over 30,000 records from the Bank dataset released by the UCI Machine Learning Repository.\footnote{\url{https://archive.ics.uci.edu/ml/datasets/bank+marketing}}
The features used are 7 including client age, campaign contact duration, economic indicators, and employment statistics.

\paragraph{HAR}
This consists of 10,299 records from the Human Activity Recognition (HAR) dataset.\footnote{\url{https://archive.ics.uci.edu/ml/datasets/human+activity+recognition+using+smartphones}}
The features, originally 561, are reduced to 10 using PCA, and the activity labels (e.g., walking, sitting, standing) are used as ground truth for clustering into 6 clusters.

\paragraph{S1}
This consists of 5,000 records from the S1 dataset, a synthetic dataset created by Fränti and Virmajoki~\cite{franti2006iterative}.\footnote{\url{https://cs.joensuu.fi/sipu/datasets/}}
The features are 2, representing two-dimensional points, which are generated as Gaussian clusters around 15 given centroids.
This dataset is also used in~\cite{su2016differentially, mohassel2020practical}.

\vspace{5pt}

In addition, we generate synthetic datasets on our own in the same style of S1 to assess our approach on different combinations of data points, clusters, and dimensions.
A summary of all datasets we use is found in Table~\ref{tab:datasets}.

\begin{table}
  \caption{List of datasets.}
  \label{tab:datasets}
  \begin{tabular}{cccc}
    \toprule
    Dataset & Points ($\numPoints$) & Clusters ($\numClusters$) & Dimensions ($\numDims$) \\
    \midrule
    Bank & 30090 & 7 & 7 \\
    HAR & 10299 & 6 & 10 \\
    Loan & 60000 & 5 & 16 \\
    Taxi & 100000 & 5 & 8 \\
    S1 & 5000 & 15 & 2 \\
    Synth-\numPoints-\numClusters-\numDims & \numPoints & \numClusters & \numDims \\
  \bottomrule
\end{tabular}
\end{table}

\subsection{Runtime}

\begin{table*}
  \caption{Runtime comparison with state-of-the-art for 10 iterations on a variety of synthetic datasets.}
  \label{tab:comparison_mohassel}
  \begin{subtable}[b]{\textwidth}
  \centering
  \caption{Runtime of Mohassel, Rosulek, and Trieu~\cite{mohassel2020practical}.}
  \begin{tabular}{%
        l|%
        >{\raggedleft\arraybackslash}p{25pt}%
        >{\raggedleft\arraybackslash}p{25pt}%
        >{\raggedleft\arraybackslash}p{25pt}|%
        >{\raggedleft\arraybackslash}p{25pt}%
        >{\raggedleft\arraybackslash}p{25pt}%
        >{\raggedleft\arraybackslash}p{25pt}|%
        >{\raggedleft\arraybackslash}p{25pt}%
        >{\raggedleft\arraybackslash}p{25pt}%
        >{\raggedleft\arraybackslash}p{25pt}|%
        >{\raggedleft\arraybackslash}p{25pt}%
        >{\raggedleft\arraybackslash}p{25pt}%
        >{\raggedleft\arraybackslash}p{25pt}%
        }
    \toprule
    Network
    & \multicolumn{3}{l|}{$n = 1000$}
    & \multicolumn{3}{l|}{$n = 10000$}
    & \multicolumn{3}{l|}{$n = 100000$}
    & \multicolumn{3}{l}{$n = 1000000$}
    \\
    Configuration
    & $k = 2$
    & $k = 5$
    & $k = 8$
    & $k = 2$
    & $k = 5$
    & $k = 8$
    & $k = 2$
    & $k = 5$
    & $k = 8$
    & $k = 2$
    & $k = 5$
    & $k = 8$
    \\
    \midrule
    LAN10000    & 15.94s & 56.00s & 1.60m & 2.96m & 9.91m & 16.95m & 29.91m & 1.66h & 2.84h & 4.99h & 16.62h & 1.18d \\
    LAN1000     & 18.31s & 1.13m & 1.94m & 3.47m & 11.67m & 21.14m & 35.10m & 1.95h & 3.55h & 5.86h & 19.52h & 1.48d \\
    LAN500      & 27.53s & 1.66m & 2.85m & 4.69m & 16.94m & 28.33m & 46.98m & 2.83h & 4.72h & 7.83h & 1.18d & 1.97d \\
    regWAN500   & 2.05m & 8.04m & 14.05m & 20.40m & 1.32h & 2.31h & 3.40h & 13.20h & 23.10h & 1.42d & 5.50d & 9.63d \\
    regWAN250   & 2.89m & 11.35m & 19.87m & 29.03m & 1.88h & 3.30h & 4.84h & 18.84h & 1.37d & 2.02d & 7.85d & 13.74d \\
    regWAN100   & 3.76m & 14.85m & 25.70m & 37.49m & 2.45h & 4.29h & 6.25h & 1.02d & 1.79d & 2.60d & 10.21d & 17.89d \\
    crpWAN500   & 8.79m & 34.77m & 1.02h & 1.46h & 5.79h & 10.13h & 14.54h & 2.41d & 4.22d & 6.06d & 24.11d & 42.17d \\
    ccWAN200    & 17.32m & 1.14h & 2.00h & 2.86h & 11.40h & 19.93h & 1.19d & 4.75d & 8.30d & 11.89d & 47.50d & 82.98d \\
    ccWAN100    & 20.70m & 1.37h & 2.40h & 3.43h & 13.67h & 23.90h & 1.43d & 5.69d & 9.95d & 14.28d & 56.94d & 99.53d \\
    ccWAN50     & 25.87m & 1.73h & 3.02h & 4.29h & 17.11h & 1.25d & 1.79d & 7.12d & 12.49d & 17.88d & 71.19d & 124.9d \\
    \midrule
    Comm. Size  & 230MB & 910MB & 1.59GB & 2.25GB & 9.00GB & 15.7GB & 22.4GB & 89.9GB & 157GB & 224GB & 899GB & 1.57TB \\
  \bottomrule
\end{tabular}
\end{subtable}
\\[1em]
\begin{subtable}[b]{\textwidth}
  \centering
  \caption{Runtime and speedup of our solution.}
  \begin{tabular}{%
        l|%
        >{\raggedleft\arraybackslash}p{25pt}%
        >{\raggedleft\arraybackslash}p{25pt}%
        >{\raggedleft\arraybackslash}p{25pt}|%
        >{\raggedleft\arraybackslash}p{25pt}%
        >{\raggedleft\arraybackslash}p{25pt}%
        >{\raggedleft\arraybackslash}p{25pt}|%
        >{\raggedleft\arraybackslash}p{25pt}%
        >{\raggedleft\arraybackslash}p{25pt}%
        >{\raggedleft\arraybackslash}p{25pt}|%
        >{\raggedleft\arraybackslash}p{25pt}%
        >{\raggedleft\arraybackslash}p{25pt}%
        >{\raggedleft\arraybackslash}p{25pt}%
        }
    \toprule
    Network
    & \multicolumn{3}{l|}{$n = 1000$}
    & \multicolumn{3}{l|}{$n = 10000$}
    & \multicolumn{3}{l|}{$n = 100000$}
    & \multicolumn{3}{l}{$n = 1000000$}
    \\
    Configuration
    & $k = 2$
    & $k = 5$
    & $k = 8$
    & $k = 2$
    & $k = 5$
    & $k = 8$
    & $k = 2$
    & $k = 5$
    & $k = 8$
    & $k = 2$
    & $k = 5$
    & $k = 8$
    \\
    \midrule
    LAN10000    & 52.44s & 57.38s & 1.03m & 55.81s & 1.70m & 2.08m & 1.41m & 2.51m & 5.84m & 1.94m & 22.21m & 1.04h \\[-4pt]
                & \footnotesize\textcolor{red}{0.30x} & \footnotesize\textcolor{red}{0.98x} & \footnotesize\textcolor{supergreen}{1.55x} & \footnotesize\textcolor{supergreen}{3.18x} & \footnotesize\textcolor{supergreen}{5.83x} & \footnotesize\textcolor{supergreen}{8.15x} & \footnotesize\textcolor{supergreen}{21.2x} & \footnotesize\textcolor{supergreen}{39.7x} & \footnotesize\textcolor{supergreen}{29.2x} & \footnotesize\textcolor{supergreen}{154x} & \footnotesize\textcolor{supergreen}{44.9x} & \footnotesize\textcolor{supergreen}{27.2x} \\
    LAN1000     & 51.88s & 57.04s & 1.03m & 55.76s & 1.70m & 2.08m & 1.41m & 2.53m & 5.86m & 1.98m & 22.57m & 1.06h \\[-4pt]
                & \footnotesize\textcolor{red}{0.35x} & \footnotesize\textcolor{supergreen}{1.19x} & \footnotesize\textcolor{supergreen}{1.87x} & \footnotesize\textcolor{supergreen}{3.73x} & \footnotesize\textcolor{supergreen}{6.88x} & \footnotesize\textcolor{supergreen}{10.2x} & \footnotesize\textcolor{supergreen}{24.9x} & \footnotesize\textcolor{supergreen}{46.2x} & \footnotesize\textcolor{supergreen}{36.4x} & \footnotesize\textcolor{supergreen}{177x} & \footnotesize\textcolor{supergreen}{51.9x} & \footnotesize\textcolor{supergreen}{33.5x} \\
    LAN500      & 52.79s & 56.82s & 1.03m & 56.09s & 1.73m & 2.08m & 1.42m & 2.53m & 5.89m & 2.04m & 22.53m & 1.06h \\[-4pt]
                & \footnotesize\textcolor{red}{0.52x} & \footnotesize\textcolor{supergreen}{1.76x} & \footnotesize\textcolor{supergreen}{2.77x} & \footnotesize\textcolor{supergreen}{5.01x} & \footnotesize\textcolor{supergreen}{9.82x} & \footnotesize\textcolor{supergreen}{13.6x} & \footnotesize\textcolor{supergreen}{33.1x} & \footnotesize\textcolor{supergreen}{67.0x} & \footnotesize\textcolor{supergreen}{48.1x} & \footnotesize\textcolor{supergreen}{231x} & \footnotesize\textcolor{supergreen}{75.4x} & \footnotesize\textcolor{supergreen}{44.6x} \\
    regWAN500   & 52.73s & 56.99s & 1.03m & 56.61s & 1.70m & 2.09m & 1.44m & 2.57m & 5.93m & 2.13m & 22.11m & 1.04h \\[-4pt]
                & \footnotesize\textcolor{supergreen}{2.33x} & \footnotesize\textcolor{supergreen}{8.46x} & \footnotesize\textcolor{supergreen}{13.6x} & \footnotesize\textcolor{supergreen}{21.6x} & \footnotesize\textcolor{supergreen}{46.5x} & \footnotesize\textcolor{supergreen}{66.4x} & \footnotesize\textcolor{supergreen}{142x} & \footnotesize\textcolor{supergreen}{308x} & \footnotesize\textcolor{supergreen}{234x} & \footnotesize\textcolor{supergreen}{957x} & \footnotesize\textcolor{supergreen}{358x} & \footnotesize\textcolor{supergreen}{222x} \\
    regWAN250   & 53.49s & 57.33s & 1.04m & 56.30s & 1.72m & 2.10m & 1.48m & 2.62m & 6.02m & 2.43m & 23.23m & 1.09h \\[-4pt]
                & \footnotesize\textcolor{supergreen}{3.24x} & \footnotesize\textcolor{supergreen}{11.9x} & \footnotesize\textcolor{supergreen}{19.1x} & \footnotesize\textcolor{supergreen}{30.9x} & \footnotesize\textcolor{supergreen}{65.8x} & \footnotesize\textcolor{supergreen}{94.2x} & \footnotesize\textcolor{supergreen}{197x} & \footnotesize\textcolor{supergreen}{432x} & \footnotesize\textcolor{supergreen}{329x} & \footnotesize\textcolor{supergreen}{1196x} & \footnotesize\textcolor{supergreen}{486x} & \footnotesize\textcolor{supergreen}{302x} \\
    regWAN100   & 53.84s & 58.36s & 1.05m & 57.64s & 1.73m & 2.11m & 1.54m & 2.70m & 6.00m & 2.99m & 22.68m & 1.07h \\[-4pt]
                & \footnotesize\textcolor{supergreen}{4.19x} & \footnotesize\textcolor{supergreen}{15.3x} & \footnotesize\textcolor{supergreen}{24.5x} & \footnotesize\textcolor{supergreen}{39.0x} & \footnotesize\textcolor{supergreen}{84.9x} & \footnotesize\textcolor{supergreen}{122x} & \footnotesize\textcolor{supergreen}{244x} & \footnotesize\textcolor{supergreen}{545x} & \footnotesize\textcolor{supergreen}{429x} & \footnotesize\textcolor{supergreen}{1252x} & \footnotesize\textcolor{supergreen}{648x} & \footnotesize\textcolor{supergreen}{403x} \\
    crpWAN500   & 53.60s & 58.14s & 1.05m & 57.46s & 1.74m & 2.11m & 1.45m & 2.59m & 6.07m & 2.12m & 22.24m & 1.05h \\[-4pt]
                & \footnotesize\textcolor{supergreen}{9.84x} & \footnotesize\textcolor{supergreen}{35.9x} & \footnotesize\textcolor{supergreen}{57.9x} & \footnotesize\textcolor{supergreen}{91.2x} & \footnotesize\textcolor{supergreen}{200x} & \footnotesize\textcolor{supergreen}{287x} & \footnotesize\textcolor{supergreen}{602x} & \footnotesize\textcolor{supergreen}{1340x} & \footnotesize\textcolor{supergreen}{1000x} & \footnotesize\textcolor{supergreen}{4117x} & \footnotesize\textcolor{supergreen}{1561x} & \footnotesize\textcolor{supergreen}{969x} \\
    ccWAN200    & 56.49s & 1.00m & 1.09m & 59.04s & 1.77m & 2.16m & 1.52m & 2.65m & 5.98m & 2.39m & 21.96m & 1.03h \\[-4pt]
                & \footnotesize\textcolor{supergreen}{18.4x} & \footnotesize\textcolor{supergreen}{68.4x} & \footnotesize\textcolor{supergreen}{111x} & \footnotesize\textcolor{supergreen}{174x} & \footnotesize\textcolor{supergreen}{387x} & \footnotesize\textcolor{supergreen}{554x} & \footnotesize\textcolor{supergreen}{1123x} & \footnotesize\textcolor{supergreen}{2579x} & \footnotesize\textcolor{supergreen}{1999x} & \footnotesize\textcolor{supergreen}{7158x} & \footnotesize\textcolor{supergreen}{3115x} & \footnotesize\textcolor{supergreen}{1930x} \\
    ccWAN100    & 56.58s & 1.02m & 1.10m & 59.30s & 1.80m & 2.20m & 1.57m & 2.71m & 6.06m & 2.78m & 22.88m & 1.07h \\[-4pt]
                & \footnotesize\textcolor{supergreen}{22.0x} & \footnotesize\textcolor{supergreen}{80.8x} & \footnotesize\textcolor{supergreen}{131x} & \footnotesize\textcolor{supergreen}{208x} & \footnotesize\textcolor{supergreen}{456x} & \footnotesize\textcolor{supergreen}{651x} & \footnotesize\textcolor{supergreen}{1308x} & \footnotesize\textcolor{supergreen}{3030x} & \footnotesize\textcolor{supergreen}{2366x} & \footnotesize\textcolor{supergreen}{7395x} & \footnotesize\textcolor{supergreen}{3584x} & \footnotesize\textcolor{supergreen}{2222x} \\
    ccWAN50     & 58.91s & 1.05m & 1.15m & 1.04m & 1.83m & 2.25m & 1.73m & 2.96m & 6.39m & 4.76m & 24.94m & 1.17h \\[-4pt]
                & \footnotesize\textcolor{supergreen}{26.4x} & \footnotesize\textcolor{supergreen}{99.0x} & \footnotesize\textcolor{supergreen}{157x} & \footnotesize\textcolor{supergreen}{248x} & \footnotesize\textcolor{supergreen}{561x} & \footnotesize\textcolor{supergreen}{801x} & \footnotesize\textcolor{supergreen}{1488x} & \footnotesize\textcolor{supergreen}{3464x} & \footnotesize\textcolor{supergreen}{2813x} & \footnotesize\textcolor{supergreen}{5405x} & \footnotesize\textcolor{supergreen}{4110x} & \footnotesize\textcolor{supergreen}{2557x} \\
    \midrule
    Comm. Size  & 17.9MB & 19.4MB & 20.0MB & 17.9MB & 19.4MB & 20.0MB & 61.9MB & 72.9MB & 76.6MB & 466MB & 563MB & 596MB \\[-4pt]
                & \footnotesize\textcolor{supergreen}{12.9x} & \footnotesize\textcolor{supergreen}{46.8x} & \footnotesize\textcolor{supergreen}{79.7x} & \footnotesize\textcolor{supergreen}{126x} & \footnotesize\textcolor{supergreen}{463x} & \footnotesize\textcolor{supergreen}{789x} & \footnotesize\textcolor{supergreen}{363x} & \footnotesize\textcolor{supergreen}{1233x} & \footnotesize\textcolor{supergreen}{2053x} & \footnotesize\textcolor{supergreen}{482x} & \footnotesize\textcolor{supergreen}{1596x} & \footnotesize\textcolor{supergreen}{2639x} \\
  \bottomrule
\end{tabular}
\end{subtable}
\end{table*}

First, we evaluate the scalability of our approach with respect to the number of data points and clusters using two-dimensional synthetic data.
Experiments are conducted on datasets of size 1,000, 10,000, 100,000, and 1,000,000 points, with 2, 5, and 8 clusters, under various network configurations.
We compare our runtime to the online phase of Mohassel, Rosulek, and Trieu~\cite{mohassel2020practical} over 10 iterations of Lloyd's algorithm.
Due to the duration of certain experiments, some runtime results are based on estimations.

The runtimes for both approaches are reported in Table~\ref{tab:comparison_mohassel}, where the results show that our proposed approach outperforms Mohassel et al.'s solution across nearly all tested settings.
The improvements are particularly pronounced for larger datasets, higher cluster counts, and constrained network environments, where communication efficiency is critical.
A detailed analysis follows.

\paragraph{Scalability}
While our solution performs comparably or slightly worse than~\cite{mohassel2020practical} on small datasets ($\numPoints = 1{,}000$), it achieves substantial speedups as the dataset size increases.
On LAN10000, our approach delivers an 8.15x speedup for $\numPoints = 10{,}000$, a 39.7x speedup for $\numPoints = 100{,}000$, and a 154x speedup for $\numPoints = 1{,}000{,}000$.

\paragraph{Impact of Network Configurations}
The performance gap between the two approaches grows significantly under constrained network conditions, such as regional WANs (regWAN) and corporate WANs (crpWAN).
For example, for $\numPoints = 100{,}000$ and $\numClusters = 5$, the speedup increases from 39.7x on LAN10000 to 3464x on crpWAN500.

In the more challenging cross-continental WAN (ccWAN) configurations, where bandwidth and latency are further restricted, our solution achieves even greater improvements.
For $\numPoints = 1{,}000{,}000$, our method delivers a 7395x speedup, processing the dataset in minutes compared to the days required by~\cite{mohassel2020practical}.

\paragraph{Efficiency for Larger Clusters ($k$)}
While the runtime of~\cite{mohassel2020practical} increases significantly with the number of clusters $k$, our solution scales more gracefully.
For instance, on regWAN500 with $\numPoints = 10{,}000$, increasing $\numClusters$ from 2 to 8 results in a 6.79x runtime increase for Mohassel et al., compared to only a 2.22x increase for our approach.
This reduced sensitivity of our solution to $\numClusters$ is particularly evident for smaller datasets.

\paragraph{Communication Overhead}
The gap in runtime between the two approaches can mainly be attributed to differences in communication costs.
The communication size of~\cite{mohassel2020practical} grows significantly with both $\numPoints$ and $\numClusters$, reaching up to 1.57 TB for $\numPoints = 1{,}000{,}000$ and $\numClusters = 8$.
Our approach, by contrast, reduces the total communication to 596 MB in the same setting.
This significant reduction favors our approach especially in networks with limited bandwidth.

\vspace{5pt}

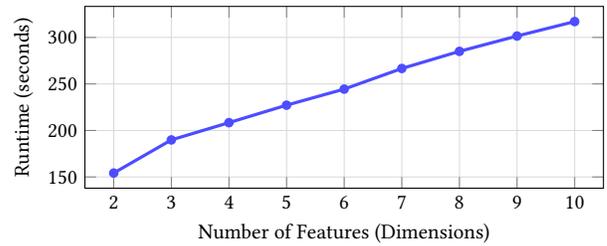
\begin{figure}
  \centering
  \begin{tikzpicture}
  \small
    \begin{axis}[
        xlabel={Number of Features (Dimensions)},
        ylabel={Runtime (seconds)},
        xmin=1.5, xmax=10.5,
        xtick={2,3,4,5,6,7,8,9,10},
        grid=both,
        major grid style={line width=.2pt, draw=gray!30},
        minor grid style={line width=.1pt, draw=gray!10},
        every axis plot/.append style={very thick,mark size=1.2pt, mark=*},
        height=4cm,
        width=\columnwidth
    ]
    
    \addplot[color=blue!70] coordinates {
        (2, 154.20)
        (3, 189.81)
        (4, 208.38)
        (5, 227.15)
        (6, 244.43)
        (7, 266.59)
        (8, 285.01)
        (9, 301.48)
        (10, 317.04)
    };
    \end{axis}
\end{tikzpicture}
  \caption{Runtime of 10 iterations of our approach for increasing dimensionality. The number of points and clusters are fixed to $\numPoints = 100{,}000$ and $\numClusters = 5$, respectively. Experiments are conducted in the regWAN500 network environment.}
  \label{fig:runtime-dimensions}
\end{figure}

We also evaluate the scalability of our solution with respect to the dimensionality of the dataset.
To achieve this, we generate 100,000 points clustered around five centroids in an increasingly higher-dimensional space.
In our setup, Alice receives one component of each point, while Bob holds all the remaining components.
Figure~\ref{fig:runtime-dimensions} presents the runtime results, which show that our solution scales linearly with the number of features in the dataset.
It is worth noting that this configuration is runtime-equivalent to scenarios where features are distributed across multiple parties.
The performance of our protocol remains largely agnostic to whether the data is divided between two parties (with one feature each) or a single party (with two features).
Consequently, Figure~\ref{fig:runtime-dimensions} can also be interpreted as reporting the runtime of our protocol in multi-party settings.

Additionally, we assess our solution on real-world datasets and compare it against the DP-based approach of Li, Wang, and Li~\cite{li2022differentially}, as shown in Table~\ref{tab:comparison_with_li}.
While their DP-based solution outperforms ours in the two-party setting, unexpectedly it becomes slower as the number of parties increases.
For example, on the Taxi dataset, our runtime never exceeds 2.60 minutes, regardless of the number of parties (the total number of features is fixed).
In contrast, the runtime of \cite{li2022differentially} increases significantly: 1.31 minutes for 2 parties, 2.59 minutes for 4 parties, and 39.61 minutes for 8 parties.
This performance degradation appears to come from the computational overhead introduced by the central server, which requires more iterations to aggregate the local clusters and the DP membership information as the number of parties increases.
Nevertheless, it is important to note that the DP-based solution comes at a high cost of accuracy, a trade-off that we further discuss in Section~\ref{sec:accuracy}.

Finally, we point out that our approach could be made even faster by employing GPU acceleration or dedicated hardware.
For instance, the work of Özcan et al.~\cite{ozcan2024heongpu} demonstrates that various HE applications can be accelerated by up to two orders of magnitude when implemented on a GeForce RTX 4090, compared to their CPU counterparts.
Based on these findings, we estimate that clustering 1,000,000 points into 5 clusters could be completed in approximately 13 seconds.
We leave this optimization to future work.

\subsection{Accuracy}
\label{sec:accuracy}

\begin{table*}
  \caption{Comparison with Li, Wang, and Li~\cite{li2022differentially}.}
  \label{tab:comparison_with_li}
  \begin{tabular}{%
        l|%
        p{35pt}%
        p{35pt}|%
        p{35pt}%
        p{35pt}%
        p{35pt}|%
        p{35pt}%
        p{35pt}%
        p{35pt}%
        }
    \toprule
    & \multicolumn{2}{l|}{Plaintext Baseline}
    & \multicolumn{3}{l|}{Our Solution}
    & \multicolumn{3}{l}{Li, Wang, and Li~\cite{li2022differentially}}
    \\
    Dataset
    & Loss
    & Accuracy
    & Loss
    & Accuracy
    & Runtime
    & Loss
    & Accuracy
    & Runtime
    \\
    \midrule
    S1 (2 parties) & 0.00286 & 93.22\% & 0.00566 & 90.75\% & 1.32m & 0.0243 & 62.25\% & 15.0s \\
    Synth-10000-8-2 (2 parties) & 0.0107 & 81.55\% & 0.0112 & 86.67\% & 2.09m & 0.0343 & 70.31\% & 18.91s \\
    HAR (2 parties) & 0.0355 & 49.63\% & 0.0417 & 45.15\% & 1.82m & 0.0686 & 30.31\% & 13.59s \\
    HAR (5 parties) & 0.0355 & 49.63\% & 0.0413 & 45.16\% & 1.85m & 0.0966 & 29.15\% & 38.87s \\
    HAR (10 parties) & 0.0355 & 49.63\% & 0.0414 & 45.21\% & 1.89m & 0.1558 & 22.83\% & 7.93m \\
    Taxi (2 parties) & 0.056 & - & 0.059 & - & 2.57m & 0.070 & - & 1.31m \\
    Taxi (4 parties) & 0.056 & - & 0.058 & - & 2.58m & 0.079 & - & 2.59m \\
    Taxi (8 parties) & 0.056 & - & 0.060 & - & 2.60m & 0.122 & - & 39.61m \\
    Bank (2 parties) & 0.165 & - & 0.187 & - & 1.83m & 0.238 & - & 24.80s \\
    Bank (4 parties) & 0.165 & - & 0.183 & - & 1.85m & 0.250 & - & 1.08m \\
    Bank (7 parties) & 0.165 & - & 0.184 & - & 1.88m & 0.296 & - & 2.38m \\
    Loan (2 parties) & 0.550 & - & 0.556 & - & 2.03m & 0.738 & - & 52.43s \\
    Loan (4 parties) & 0.550 & - & 0.554 & - & 2.05m & 0.766 & - & 1.91m \\
    Loan (8 parties) & 0.550 & - & 0.556 & - & 2.07m & 0.785 & - & 7.06m \\
    Loan (16 parties) & 0.550 & - & 0.557 & - & 2.10m & \multicolumn{3}{l}{out-of-memory} \\
  \bottomrule
\end{tabular}
\end{table*}

While assessing the utility of our approach, we compare it to the state-of-the-art DP-based solution for the vertical setting, developed by Li, Wang, and Li~\cite{li2022differentially}.
Both approaches are evaluated on multiple datasets under identical privacy parameters, specifically $\epsilon = 1$ and $\delta = 1 / \numPoints$, as recommended in~\cite{li2022differentially}.
We use two utility metrics:
\begin{itemize}
    \item Normalized k-means loss, representing the objective minimized by Lloyd's algorithm:
    $$\qquad\quad\frac{1}{\numPoints} \sum_{i = 1}^{\numPoints}{\min_{j \in \{1, \dots, \numClusters\}}{\norm{\point_i - \centroid_j}^2}} \enspace ,$$
    \item Cluster accuracy, which measures the proportion of points clustered correctly (modulo permutation of the centroids):
    $$\qquad\quad\max_{\pi \in S_\numClusters}{|\{1, \dots, \numPoints : \argmin_{j \in \{1, \dots, \numClusters\}}{\norm{\point_i - \centroid_{\pi(j)}}} = \groundTruth_i\}| / \numPoints} \enspace ,$$
    where $\groundTruth_i$ is the ground-truth cluster index of point $\point_i$, and $S_\numClusters$ is the group of permutations of size $\numClusters$.
\end{itemize}
For most real-world datasets, ground-truth cluster labels are not available, so only k-means loss is reported.
We also include the plaintext baseline (standard Lloyd's algorithm on the plaintext joint dataset) as a reference.

The results, summarized in Table~\ref{tab:comparison_with_li}, are averaged over multiple runs, as outcomes may depend on the initial centroid selection.
Overall, our approach demonstrates higher utility compared to Li, Wang, and Li~\cite{li2022differentially}, achieving results closer to the plaintext baseline.
For example, on the synthetic S1 dataset, our k-means loss is 0.00566, much closer to the plaintext baseline (0.00286) than Li's solution (0.0243).
Similarly, our cluster accuracy is 90.75\%, significantly higher than Li's 62.25\% and aligning closely with the plaintext accuracy of 93.22\%.
This trend is consistent across datasets, indicating that our approach maintains higher utility while ensuring the same privacy guarantees.

\paragraph{Utility across Parties}
Where possible, we distribute the dataset features across different number of parties.
In such cases, we observe that our solution maintains stable utility as the number of parties increases, whereas the utility of Li's solution degrades significantly.
For instance, on the HAR dataset, our accuracy remains consistent (e.g., 45.15\% for 2 parties, 45.16\% for 5 parties, and 45.21\% for 10 parties) and close to the plaintext baseline (49.63\%).
In contrast, Li's solution shows a marked decline (e.g., 30.31\% for 2 parties, 29.15\% for 5 parties, and 22.83\% for 10 parties).
Similar trends are observed in other real-world datasets like Taxi and Bank.
For example, in the Taxi dataset, our k-means loss increases only slightly from 0.059 (2 parties) to 0.060 (8 parties), whereas Li's solution degrades significantly, from 0.070 to 0.122.
This consistency highlights the robustness of our approach in scenarios where data is distributed across many parties, ensuring reliable utility regardless of the number of participants.

\vspace{5pt}

Interestingly, there are cases where our solution outperforms even the plaintext baseline.
For instance, on the Synth-10000-8-2 dataset, our clustering accuracy is 86.67\%, better than the plaintext baseline accuracy of 81.55\%.
This phenomenon likely occurs because the differentially private noise added in our approach is acting as a regularizer, leading Lloyd's algorithm towards a different local minimum.
While such scenarios are rare, they show the potential for DP noise to enhance optimization in certain settings.

\paragraph{Explaining the Utility Gap} The higher utility of our approach compared to Li's can be attributed to differences in the noise application.
Although both methods apply the same noise scale due to identical sensitivity of the centroids, Li's solution introduces noise in more elements than we do: to all local centroids generated by each party and to the corresponding membership encodings.
In contrast, our approach applies noise only to the $\numClusters$ intermediate centroids, reducing the cumulative impact of noise on the final output.

\begin{figure}
\newcommand{\subwidth}{200pt}
  \centering
  \begin{subfigure}{1\columnwidth}
  \centering
  \begin{tikzpicture}
  \pgfplotsset{scaled y ticks=false}
  \small
    \begin{axis}[
        xlabel={Privacy Budget ($\epsilon$)},
        ylabel={Loss},
        xtick={0.5,1,2,3,4,5},
        grid=both,
        ytick={0,0.01,0.02,0.03,0.04},
        yticklabels={0,.01,.02,.03,.04},
        major grid style={line width=.2pt, draw=gray!30},
        minor grid style={line width=.1pt, draw=gray!10},
        every axis plot/.append style={very thick,mark size=1.2pt, mark=*},
        height=4cm,
        width=\subwidth,
        legend style={
            at={(0.5,1.1)},
            anchor=south,
            draw=none,
            yshift=-4pt
        },
        legend columns=2,
        legend cell align={left}
    ]
    \addplot[color=blue!70] coordinates {
        (0.5, 0.008833738)
        (1, 0.005659972)
        (2, 0.004578381)
        (3, 0.004215697)
        (4, 0.004406351)
        (5, 0.004178815)
    };
    \addplot[color=red!80] coordinates {
        (0.5, 0.042605641)
        (1, 0.024326211)
        (2, 0.021531879)
        (3, 0.022254492)
        (4, 0.021475044)
        (5, 0.0212557)
    };
    \addlegendentry{our solution}
    \addlegendentry{Li, Wang, and Li~\cite{li2022differentially}}
    \draw[thin, dashed] (axis cs:0, 0.00286) -- (axis cs:6.5, 0.00286);
    \end{axis}
  \end{tikzpicture}
  \caption{Loss.}
  \label{fig:accuracy-vs-epsilon-accuracy}
  \end{subfigure}
  \begin{subfigure}{1\columnwidth}
  \centering
  \begin{tikzpicture}
  \small
    \begin{axis}[
        xlabel={Privacy Budget ($\epsilon$)},
        ylabel={Accuracy},
        xtick={0.5,1,2,3,4,5},
        grid=both,
        ymin=0.55,
        major grid style={line width=.2pt, draw=gray!30},
        minor grid style={line width=.1pt, draw=gray!10},
        every axis plot/.append style={very thick,mark size=1.2pt, mark=*},
        height=4cm,
        width=\subwidth,
        legend style={
            at={(0.5,1.1)},
            anchor=south,
            draw=none,
            yshift=-4pt
        },
        legend columns=2,
        legend cell align={left}
    ]
    \addplot[color=blue!70] coordinates {
        (0.5, 0.8471)
        (1, 0.90748)
        (2, 0.92674)
        (3, 0.929733333)
        (4, 0.924733333)
        (5, 0.928866667)
    };
    \addplot[color=red!80] coordinates {
        (0.5, 0.613622)
        (1, 0.622494)
        (2, 0.60432)
        (3, 0.6087)
        (4, 0.60945)
        (5, 0.617344)
    };
    \addlegendentry{our solution}
    \addlegendentry{Li, Wang, and Li~\cite{li2022differentially}}
    \draw[thin, dashed] (axis cs:0, 0.00286) -- (axis cs:6.5, 0.00286);
    \draw[thin, dashed] (axis cs:0, 0.932196667) -- (axis cs:5.5, 0.932196667);
    \end{axis}
  \end{tikzpicture}
  \caption{Accuracy.}
  \label{fig:accuracy-vs-epsilon-loss}
  \end{subfigure}

  \caption{K-Means loss and accuracy of our solution vs. Li, Wang, and Li~\cite{li2022differentially} on S1, for varying privacy budget. The dashed lines represent the plaintext baseline.}
  \label{fig:accuracy-vs-epsilon}
\end{figure}
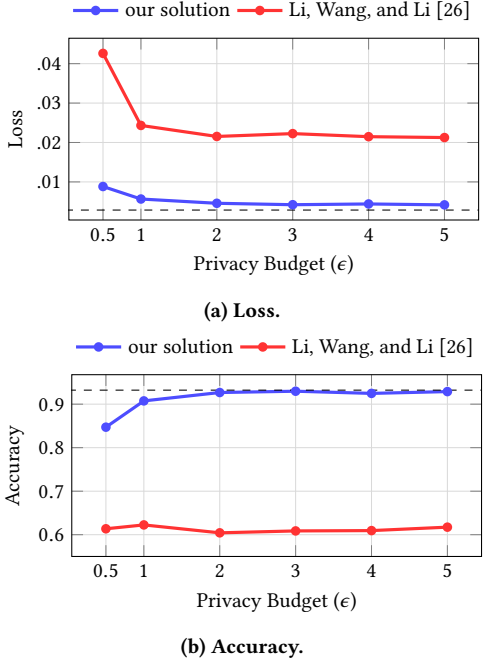

\paragraph{Privacy-Accuracy Trade-off}
Figure~\ref{fig:accuracy-vs-epsilon} shows the impact of the privacy budget ($\epsilon$) on the performance of our protocol compared to the method proposed by Li, Wang, and Li~\cite{li2022differentially}.
For $\epsilon$ ranging from 0.5 to 5, consistent with what used in their evaluation, we reported clustering accuracy and k-means loss.
Subfigure~\ref{fig:accuracy-vs-epsilon-accuracy} shows that our approach achieves lower loss for all values of $\epsilon$, with the loss stabilizing around 0.004 as $\epsilon$ increases.
In contrast, the baseline method exhibits higher and more variable loss values.
Subfigure~\ref{fig:accuracy-vs-epsilon-loss} shows that our protocol achieves higher accuracy, surpassing 90\% for $\epsilon \geq 1$, while the baseline accuracy remains below 65\% across all tested values of $\epsilon$.
These results demonstrate that our method provides better utility while maintaining strong privacy guarantees, outperforming the baseline in both loss and accuracy.

\vspace{5pt}

Finally, we point out that the noise introduced in our scheme is equivalent to what would be required to reveal only the final set of centroids.
Thus, decrypting the intermediate centroids does not require additional noise beyond what is already needed to ensure the privacy of the final output.
In other words, we use the same amount of noise that would be required for MPC solutions to provide a DP output.


\section{Related Work}

There is extensive literature on privacy-preserving k-means clustering, broadly categorized into solutions based on MPC and solutions based on DP.
We summarize key contributions in each category, highlighting their features as well as their limitations.

MPC-based solutions for collaborative clustering have been around for a few decades.
An early work by Vaidya and Clifton~\cite{vaidya2003privacy} focuses on the VP setting.
Here the authors use secret sharing to compute distances between points and centroids, then they employ three non-colluding parties to extract the closest centroid for each point, by secretly permuting the order of the clusters.
Unfortunately, their solution reveals the intermediate clusters centers.
From there other solutions followed.
Jagannathan and Wright~\cite{jagannathan2005privacy} improved on the state-of-the-art by using random sharing to compute distances and Yao's circuit evaluation~\cite{yao1986generate} to determine the minimum distance and the closest centroid, which is however still in the clear.
Bunn and Ostrovsky~\cite{bunn2007secure} managed to solve the issue of the intermediate centroids by using the Paillier partial HE scheme~\cite{paillier1999public} to compute them securely, though at the cost of a significant computational slow-down.
More recently, Mohassel, Rosulek, and Trieu~\cite{mohassel2020practical} propose a solution that hides the intermediate centroids, while keeping a relatively low runtime. To do so, the authors use a custom garbled circuit to optimize the minimum computation across shared values.
It is worth mentioning that the approaches described in \cite{jagannathan2005privacy, bunn2007secure, mohassel2020practical} are actually applicable both in the vertical- and in the horizontal-partition settings.
Other lines of work only focused on the horizontal setting, where the main challenge is the aggregation of the local centroids.
In this setting, Jha, Kruger, and McDaniel~\cite{jha2005privacy} present two solutions, one based on oblivious polynomial evaluation and the other on the Benaloh partial HE scheme~\cite{benaloh1994dense},
while Gheid and Challal~\cite{gheid2016efficient} extend Clifton’s secure sum protocol~\cite{clifton2002tools} to enable centroid aggregation without requiring strict input bounds.
Despite their moderately low computation complexity, MPC-based approaches lead to an impractical runtime for large datasets, due to their high communication costs.
Moreover, while some of these solutions manage to protect the privacy of intermediate centroids, none of them offers any privacy guarantee for the final output.

Regarding DP-based solutions, there are mainly two works focusing on collaborative k-means clustering.
In the horizontal setting, Diaa, Humphries, and Kerschbaum~\cite{diaa2024fastlloyd} bound the cluster radius and make the centroid updates relative to the previous centroids.
This approach makes the sensitivity of the updates depend on the bounded cluster radius rather than the full domain size.
The parties execute this modified version of Lloyd'd algorithm by iteratively computing and aggregating local centroid updates.
In the vertical setting, Li, Wang, and Li~\cite{li2022differentially} propose a method where each party first performs local clustering on their own features.
They then outsource a noisy version of their local clusters, along with membership information, to a central server.
This information allows the server to determine cluster assignments and compute the final clusters.
Other DP-based solutions either focus on the central setting or on the non-interactive outsource-based setting.
For the central setting, the data-owner uses a DP interactive mechanism to answer adaptively-chosen clustering queries.
In this setting we find the work of Blum et al.~\cite{blum2005practical} who, ahead of DP formalization, introduced a straightforward private adaptation of Lloyd's algorithm by injecting noise to the cluster sums and sizes (their solution will be often referred as DPLloyd).
Balcan et al.~\cite{balcan2017differentially} subsequently improved the approach utility by using the Johnson-Lindenstrauss transform~\cite{johnson1984extensions} to project the input data into a lower-dimensional space, where a small set of good candidate centroids is built. Discrete clustering is then used to find the final k centroids, which are mapped back to the original space by noisy averaging.
On the other hand, in the non-interactive setting, a private synopsis of the data is created and can be used for outsourcing computations to an untrusted server.
In this context we only mention the work by Su et al.~\cite{su2016differentially}, where the authors divide the dataset domain into a grid and outsource the noisy count of each cell to a computing server, which will then perform the clustering (EUGkM) and then optionally run a round of DPLloyd to refine the clusters. Their work is then refined in~\cite{su2017differentially}.
While collaborative DP-based approaches are sensibly faster than MPC-based ones, reaching runtimes that are comparable with plaintext ones, the amount of noise they inject often results in substantial accuracy losses for the clustering algorithm.

Motivated by these considerations, we design a new solution based on HE.
While HE has been employed in collaborative clustering before, prior approaches show significant limitations.
For instance, Almutairi, Coenen, and Dures~\cite{almutairi2017k} attempt to solve collaborative clustering by employing Liu's HE scheme~\cite{liu2015practical} along with an updatable distance matrix to store distances between points.
However, the distance matrix is shared in plaintext, leaking information about the private input.
Moreover, it appears that Liu's scheme has been broken~\cite{wang2015notes}.
Wu et al.~\cite{wu2020secure} employ a combination of random permutations and the YASHE scheme~\cite{bos2013improved} to outsource computations to two external non-colluding parties.
However, they leak ratios of distances between points.
Jäschke and Armknecht~\cite{jaschke2018unsupervised} use TFHE~\cite{chillotti2016faster} to encrypt the input data and replace the computationally expensive division by the encrypted cluster size with a division by a constant through padding each cluster with its corresponding previous centroid.
While their work manages to avoid any intermediate privacy leakage, the computation time turns out to be impractical, with the authors reporting a runtime of 26 days to cluster a dataset of 400 two-dimensional points into 3 clusters.
Our work advances the field by employing a new method for efficiently computing the minimum under HE, delivering a privacy-preserving solution that is computationally practical, while keeping a low communication cost and preserving the accuracy of the result.


\section{Conclusion}

This paper presents a novel protocol for performing k-means clustering on vertically partitioned datasets across multiple parties, while ensuring data privacy throughout the process.
The proposed method employs an efficient approach to securely compute the argmin operation under homomorphic encryption, achieving low local computation costs while benefiting from the reduced communication overhead inherent to HE.
We evaluated the scalability and performance of the protocol, showing that it can cluster datasets with millions of points within minutes, even in constrained network settings.
These results highlight its practicality for real-world use cases.
Moreover, the proposed approach outperforms state-of-the-art solutions in both runtime efficiency and clustering accuracy, while maintaining equivalent privacy guarantees.




\begin{anonsuppress}
\begin{acks}
We thank Hao Wu for the enlightening discussions on Differential Privacy.
\end{acks}
\end{anonsuppress}


\bibliographystyle{ACM-Reference-Format}
\bibliography{bibliography}


\clearpage
\appendix

\section{Recursive Matrix Operations}
\label{app:rec_matrix_ops}

We report the pseudocode --- taken from~\cite{mazzone2024efficient} --- for $\sumR$, $\sumC$, $\replR$, $\replC$, $\transR$, $\transC$ for a square matrix with $\vectorLength$ number of rows/columns.
The matrix is assumed to be padded in such a way that $\vectorLength$ is a power of 2.

\begin{algorithm}
\caption{\sumR}
\label{alg:sumR}
\begin{algorithmic}[1]
\Require $X$ encryption of a square matrix of size $\vectorLength$.
\Ensure $X$ encryption of a row vector.
\For{$i = 0, \dots, \log{\vectorLength} - 1$}
    \State $X \gets X + (X \ll \vectorLength \cdot 2^i)$
\EndFor
\State $X \gets\maskR(X, 0)$
\State \Return $X$
\end{algorithmic}
\end{algorithm}

\begin{algorithm}
\caption{\sumC}
\label{alg:sumC}
\begin{algorithmic}[1]
\Require $X$ encryption of a square matrix of size $\vectorLength$.
\Ensure $X$ encryption of a column vector.
\For{$i = 0, \dots, \log{\vectorLength} - 1$}
    \State $X \gets X + (X \ll 2^i)$
\EndFor
\State $X \gets\maskC(X, 0)$
\State \Return $X$
\end{algorithmic}
\end{algorithm}

\begin{algorithm}
\caption{\replR}
\label{alg:replR}
\begin{algorithmic}[1]
\Require $X$ encryption of a row vector of size $\vectorLength$.
\Ensure $X$ encryption of a square matrix.
\For{$i = 0, \dots, \log{\vectorLength} - 1$}
    \State $X \gets X + (X \gg \vectorLength \cdot 2^i)$
\EndFor
\State \Return $X$
\end{algorithmic}
\end{algorithm}

\begin{algorithm}
\caption{\replC}
\label{alg:replC}
\begin{algorithmic}[1]
\Require $X$ encryption of a column vector of size $\vectorLength$.
\Ensure $X$ encryption of a square matrix.
\For{$i = 0, \dots, \log{\vectorLength} - 1$}
    \State $X \gets X + (X \gg 2^i)$
\EndFor
\State \Return $X$
\end{algorithmic}
\end{algorithm}

\begin{algorithm}
\caption{\transR}
\label{alg:transR}
\begin{algorithmic}[1]

\Require $X$ encryption of a vector $x$ encoded as a row.

\Ensure $X$ encryption of the vector $x$ encoded as a column.

\For{$i = 1, \dots, \ceil{\log{\vectorLength}}$}
    \State $X \gets X + (X \gg \vectorLength(\vectorLength-1) / 2^i)$
\EndFor
\State $X \gets \maskC(X,0)$
\State \Return $X$
\end{algorithmic}
\end{algorithm}

\begin{algorithm}
\caption{\transC}
\label{alg:transC}
\begin{algorithmic}[1]

\Require $X$ encryption of a vector $x$ encoded as a column.

\Ensure $X$ encryption of the vector $x$ encoded as a row.

\For{$i = 1, \dots, \ceil{\log{\vectorLength}}$}
    \State $X \gets X + (X \ll \vectorLength(\vectorLength-1) / 2^i)$
\EndFor
\State $X \gets \maskR(X,0)$
\State \Return $X$
\end{algorithmic}
\end{algorithm}


\section{Replication with No Padding}
\label{app:repl-no-padding}
Here we provide the pseudocode of the replication algorithm described in Section~\ref{sec:main-construction} as Algorithm~\ref{alg:super-repl}.
Note that we handle two different cases whether the starting point is in the first or second half of the vector.

\vfill\eject
\begin{algorithm}
\caption{Replication with no Padding}
\label{alg:super-repl}
\begin{algorithmic}[1]
\Require $\encVector$ encryption of $\vector = (0, \dots, 0, \vector_i, 0, \dots, 0) \in \R^\vectorLength$.
\Ensure $\encReplResult$ encryption of $(\vector_i, \dots, \vector_i) \in \R^\vectorLength$.
\State $\encReplResult \gets \encVector$
\State $\texttt{sizeLeft} \gets \numClusters - 2^{\floor{\log{\numClusters}}}$
\If{$i < \numClusters / 2$}
    \State $\texttt{adjPos} = i$
\Else
    \State $\texttt{adjPos} = i - \texttt{sizeLeft}$
\EndIf
\State $\texttt{partial} = []$
\State $\texttt{partial.append}(\encReplResult)$
\For{$l = 0, \dots, \floor{\log{\numClusters}} - 1$}
    \If{$(adjPos \gg l) \land 1$}
        \State $\encReplResult \gets \encReplResult + (\encReplResult \ll 2^l)$
    \Else
        \State $\encReplResult \gets \encReplResult + (\encReplResult \gg 2^l)$
    \EndIf
    \State \texttt{partial.append}($\encReplResult$)
\EndFor
\While{$\texttt{sizeLeft} > 0$}
    \State $\texttt{highestPow2} \gets \lfloor \log(\texttt{sizeLeft}) \rfloor$
    \If{$i < \numClusters / 2$}
        \State $\texttt{adjIndex} \gets \numClusters - \texttt{sizeLeft}$ \newline
        \hspace*{8em} $+\, (\texttt{adjPos} \land (2^{\texttt{highestPow2}} - 1)) - i$
        \State $\encReplResult \gets \encReplResult + (\texttt{partial}[\texttt{highestPow2}] \gg \texttt{adjIndex})$
    \Else
        \State $\texttt{adjIndex} \gets i - (\texttt{adjPos} \land (2^{\texttt{highestPow2}} - 1))$ \newline
        \hspace*{8em} $-\, \texttt{sizeLeft} + 2^{\texttt{highestPow2}}$
        \State $\encReplResult \gets \encReplResult + (\texttt{partial}[\texttt{highestPow2}] \ll \texttt{adjIndex})$
    \EndIf
    \State $\texttt{sizeLeft} \gets \texttt{sizeLeft} - 2^{\texttt{highestPow2}}$
\EndWhile
\State \Return $\encReplResult$
\end{algorithmic}
\end{algorithm}


\end{document}